\documentclass{article}

\usepackage{microtype}
\usepackage{graphicx}
\usepackage{subfigure}
\usepackage{booktabs} 
\usepackage{paralist}
\usepackage{hyperref}
\usepackage{tablefootnote}

\newcommand{\ul}[1]{{\underline{#1}}}

\newcommand{\chiu}[1]{\textcolor{teal}{[Keyue: #1 ]}}

\usepackage[accepted]{icml2024}

\usepackage{amsmath}
\usepackage{amssymb}
\usepackage{mathtools}
\usepackage{amsthm}

\usepackage[capitalize,noabbrev]{cleveref}
\usepackage{multirow}

\theoremstyle{plain}
\newtheorem{theorem}{Theorem}[section]
\newtheorem{proposition}[theorem]{Proposition}

\theoremstyle{definition}

\theoremstyle{remark}

\usepackage[textsize=tiny]{todonotes}
\usepackage{subcaption}
\usepackage{lineno}

\newcommand{\bx}{\mathbf{x}}
\newcommand{\bv}{\mathbf{v}}
\newcommand{\bbR}{\mathbb{R}}
\newcommand{\calP}{\mathcal{P}}
\newcommand{\calM}{\mathcal{M}}

\newcommand{\bR}{\mathbf{R}}
\newcommand{\bt}{\mathbf{t}}
\newcommand{\bbE}{\mathbb{E}}
\newcommand{\by}{\mathbf{y}}
\newcommand{\bm}{\mathbf{m}}
\newcommand{\bp}{\mathbf{p}}
\newcommand{\be}{\mathbf{e}}
\newcommand{\bh}{\mathbf{h}}
\newcommand{\btheta}{\boldsymbol{\theta}}
\newcommand{\bPhi}{\boldsymbol{\Phi}}
\newcommand{\bmu}{\boldsymbol{\mu}}
\newcommand{\bb}{\mathbf{b}}
\newcommand{\beps}{\boldsymbol{\epsilon}}
\newcommand{\bQ}{\mathbf{Q}}

\newcommand{\defeq}{\stackrel{\text{def}}{:=}}

\newcommand{\eg}{\emph{e.g.}}
\newcommand{\ie}{\emph{i.e.}}
\newcommand{\aka}{\emph{a.k.a}}

\newcommand{\kevin}[1]{{\color{blue}[qu: #1]}}
\newcommand{\mycomment}[1]{}

\newcommand{\ours}{MolCRAFT}

\icmltitlerunning{MolCRAFT: Structure-Based Drug Design in Continuous Parameter Space}

\begin{document}

\twocolumn[
\icmltitle{
MolCRAFT: Structure-Based Drug Design in Continuous Parameter Space
}



\icmlsetsymbol{equal}{*}

\begin{icmlauthorlist}
\icmlauthor{Yanru Qu}{equal,air,uiuc}
\icmlauthor{Keyue Qiu}{equal,air,thu}
\icmlauthor{Yuxuan Song}{equal,air,thu}
\icmlauthor{Jingjing Gong}{air} \\
\icmlauthor{Jiawei Han}{uiuc}
\icmlauthor{Mingyue Zheng}{shanghai}
\icmlauthor{Hao Zhou}{air}
\icmlauthor{Wei-Ying Ma}{air}
\end{icmlauthorlist}

\icmlaffiliation{uiuc}{University of Illinois Urbana-Champaign, USA}
\icmlaffiliation{thu}{Department of Computer Science and Technology, Tsinghua University}
\icmlaffiliation{air}{Institute for AI Industry Research (AIR), Tsinghua University}
\icmlaffiliation{shanghai}{Shanghai Institute of Materia Medica, Chinese Academy of Sciences}

\icmlcorrespondingauthor{Hao Zhou}{zhouhao@air.tsinghua.edu.cn}
\icmlcorrespondingauthor{Jingjing Gong}{gongjingjing@air.tsinghua.edu.cn}

\icmlkeywords{Bayesian Flow Network, Structure-based Drug Design}

\vskip 0.3in
]



\printAffiliationsAndNotice{\icmlEqualContribution{Work was done during Yanru Qu's internship at Tsinghua AIR.}} 

\begin{abstract}
Generative models for structure-based drug design (SBDD) have shown promising results in recent years. 
Existing works mainly focus on how to generate molecules with higher binding affinity, ignoring the feasibility prerequisites for generated 3D poses and resulting in \emph{false positives}. We conduct thorough studies on key factors of ill-conformational problems when applying autoregressive methods and diffusion to SBDD, including mode collapse and hybrid continuous-discrete space. 
In this paper, we introduce MolCRAFT, the first SBDD model that operates in the continuous parameter space, together with a novel noise reduced sampling strategy. 
Empirical results show that our model consistently achieves superior performance in binding affinity with more stable 3D structure, demonstrating our ability to accurately model interatomic interactions. To our best knowledge, MolCRAFT is the first to achieve reference-level Vina Scores (-6.59 kcal/mol) with comparable molecular size, outperforming other strong baselines
by a wide margin (-0.84 kcal/mol).
Code is available at \url{https://github.com/AlgoMole/MolCRAFT}.
\end{abstract}

\section{Introduction}




Structure-based drug design (SBDD) advances drug discovery by leveraging 3D structures of biological targets, thereby facilitating efficient and rational design of molecules within a certain chemical space of interests \citep{wang_deep_2022, isert_structure-based_2023}. 
Recently, generative models for molecules have emerged as a promising direction, which could streamline SBDD by directly proposing desired molecules, eliminating the need for exhaustive blind search in the vast space \cite{doi:10.1021/acs.jmedchem.8b01048, luo_3d_2022}.
Recent progress in SBDD can be divided into two categories, \ie~auto-regressive models \cite{luo_3d_2022, peng_pocket2mol_2022, zhang_molecule_2023} as next-token prediction for text generation, and diffusion models \cite{guan_3d_2023, guan_decompdiff_2023} as for image generation.

The essential criteria for drug-like candidate molecules are outlined as follows: (i) \emph{high affinity} towards specific binding sites (\emph{a.k.a}, protein pockets), where a higher affinity indicates better performance, (ii) \emph{satisfactory drug-like properties}, such as synthesizability and drug-likeness scores, which often serve as thresholds for filtering out unfavorable compounds \citep{ursu2011understanding, tian2015application}, and (iii) \emph{well-conformational 3D structure}, which needs special attention as there is a danger of generating unrealistic 3D conformations yet with deceptively high affinities. 

However, current generative models focus primarily on (i) and (ii), whereas we observe that the generated molecules often fail to meet all criteria simultaneously, especially for (iii) conformational stability.
This challenge manifests as the \textit{False Positives} phenomenon~(FP) in generative modeling of SBDD, where models yield molecules that reside outside the true molecular manifold yet appear to exhibit good binding affinity after redocking.
Specifically, these molecules suffer from \emph{distorted structure}, displaying problematically unusual topology, and \emph{inferior binding mode}, whereby the generated poses fail to capture true interactions and may even violate biophysical constraints, and thus go through post-fixes and significant rearrangements from docking software.
Such problems threaten to jeopardize reliable model assessment, ultimately hindering their application in SBDD (Sec.~\ref{subsec:failure_modes}).

Both autoregressive and diffusion-based models exhibit challenges with generating accurate molecular conformations, yet these issues stem from distinct causes. In Sec.~\ref{subsec:mode_collapse}, we delve into the \emph{mode collapse} issue faced by autoregressive methods. Empirically, they tend to repeatedly generate a limited number of specific (sub-)structures due to an unnatural atom ordering imposed during generation. 
On the other hand, the problem with diffusion-based models is attributed to \emph{denoising in hybrid yet highly twisted space}, which is a blend of discrete atomic types and continuous atomic coordinates. 
Different modalities need to be carefully handled in the hybrid space, 
and lack of consideration might result in severely strained and infeasible outputs~(Sec.~\ref{subsec:hybrid_space}).

\begin{figure*}[ht]
    \centering
    \includegraphics[width=\textwidth]{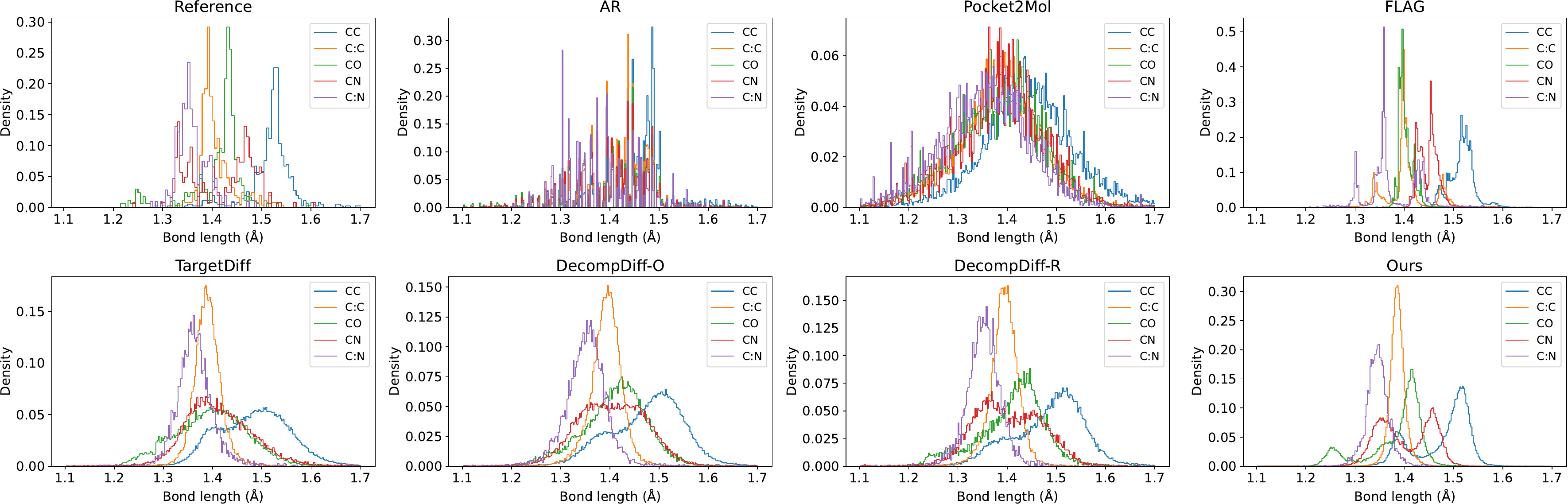}
    \caption{Bond length distribution of reference and generated molecules by autoregressive models (upper row) and non-autoregressive models (lower row) for top-5 frequent bond types.}
    \label{fig:all_len}
\end{figure*}

Notably, DecompDiff~\cite{guan_decompdiff_2023} proposes to inject the molecular inductive bias by manually decomposing ligands into arms and scaffolds priors before training, and utilizing validity guidance in sampling.
However, it cannot fully address the ill-conformational problem, since the inductive bias is simply impossible to enumerate.
As shown in Fig.~\ref{fig:all_len}, for common C-N and C-O bond with two modes of typical length distribution, nearly all SBDD models are struggling to fit this substructural pattern. More visualization results can be referred to in Fig.~\ref{fig:other_len}, \ref{fig:other_angle}, \ref{fig:other_torsion}, Appendix~\ref{sec:exp_detail}.

\begin{figure}[t]
\centering
\includegraphics[width=0.9\linewidth]{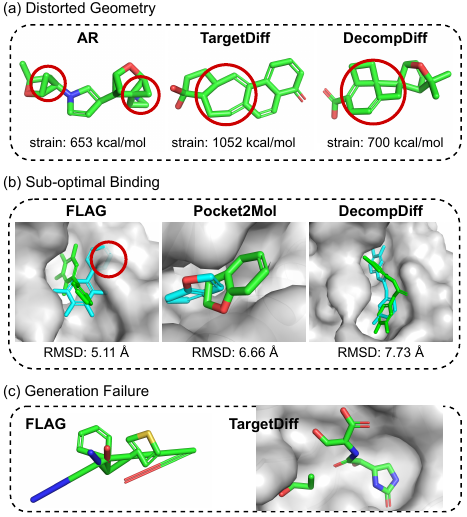}
\caption{Typical failure modes. 
(\textbf{a}) Unusual 3-membered rings generated by AR, large fused rings with more than 7 atoms generated by diffusion models. 
(\textbf{b}) Examples of steric clashes by FLAG, and other ligand undergoing significant conformational rearrangements upon redocking (Before: {\color{cyan}{blue}}. After: {\color{green}{green}}).
(\textbf{c}) Failures in generation process. Left: atoms mis-connected in autoregressive sampling. Right: incomplete molecules with multiple components.}
\label{fig:failure_mode}
\vspace{-15pt}

\end{figure}

In order to capture the complicated data manifold for molecules, we take a shift to a unified continuous parameter space instead of a hybrid space, inspired by \citet{graves2023bayesian}. We propose \ours~(\ul{C}ontinuous pa\ul{RA}meter space \ul{F}acilitated molecular genera\ul{T}ion), which not only alleviates the mode collapse issue by non-autoregressive generation as in its diffusion counterparts, but also addresses the continuous-discrete gap by applying continuous noise and smooth transformation.
Our contributions can be summarized as follows:
\begin{itemize}
\item We investigate the challenges of current SBDD models, and identify key issues including the mode collapse of autoregressive methods, and the gap of continuous-discrete space when applying diffusion models.
\item We propose \ours~to address these two issues, which is a unified SE-(3) equivariant generative model, equipped with sampling in the parameter space that avoids further noise. 
\item We conduct comprehensive evaluation under controlled molecular sizes. Experiments show that our model generates high-affinity binders with feasible 3D poses. To our best knowledge, we are the first to achieve \textbf{reference-level Vina Scores} (-6.59 vs. -6.36 kcal/mol) with comparable molecule size, outperforming other strong baselines by a wide margin (-0.84 kcal/mol).
\end{itemize}

\section{Challenges of Generative Models in SBDD}\label{sec:challenge}
We provide an overview of current obstacles in pocket-based generation.
We summarize common failures in Sec.~\ref{subsec:failure_modes}, and then investigate the underlying problems, \emph{i.e.} the \emph{mode collapse} issue of autoregressive-based models in Sec.~\ref{subsec:mode_collapse}, and \emph{hybrid denoising} issue of diffusion-based models in Sec.~\ref{subsec:hybrid_space}. Based on the aforementioned challenges, we propose to generate molecules in the continuous parameter space.

\subsection{Failure Modes of Generated Molecules}\label{subsec:failure_modes}
As shown in Fig.~\ref{fig:failure_mode}, we divide undesired molecules in SBDD into three categories:
\begin{itemize}
    \item[(a)] \textbf{Distorted geometry}. We visualize the generated molecules at median strain energy (see Table~\ref{tab:main_results}), and models tend to produce either too many uncommon 3- or 4-membered rings, or extra-large rings with unstable structures, leading to much higher strain energy.
    \item[(b)] \textbf{Inferior binding mode}. We observe a notable number of generated ligand conformations rearrange drastically after redocking, with some even violating biophysical constraints and producing steric clashes with the protein surface. This suggests that 3D SBDD models do not capture true interatomic interactions and rely on post-fixing via redocking as noted by \citet{harris2023benchmarking}, which severely harms the credibility of generating molecules directly in 3D space.
    \item[(c)] \textbf{Generation failure}. Autoregressive models tend to misplace an element and terminate prematurely, while diffusion models might generate incomplete molecules with disconnected parts, limiting sample efficiency.
\end{itemize}




The above problems hinder the applicability of SBDD models. In the following sections, we provide deeper understanding of the problematic methods underlying these failures.

\subsection{Molecular Mode Collapse}\label{subsec:mode_collapse}
The \emph{mode collapse} issue focuses on the empirical performance of SBDD methods that tend to generate a limited number of specific (sub-)structures, where atom-based autoregressive models have displayed a particular preference for certain modes.
We provide quantitative results from both the chemical and geometrical perspectives.


Chemical assessment is shown in Table~\ref{tab:chemical_mode_collapse}. 
In order to measure \emph{molecular distribution}, we report the percentage of unique samples ({\bf Unique}) averaged on different pockets.\footnote{Here we remove all post-filters from autoregressive models that avoid generating duplicate or invalid molecules, in order to faithfully demonstrate their performances. In all other experiments, we stick to the original implementation.} It can be seen that the ratio of unique molecules of AR \citep{luo_3d_2022} and Pocket2Mol \citep{peng_pocket2mol_2022} is considerably lower than other counterparts. Moreover, DecompDiff \citep{guan_decompdiff_2023} is also found to generate repeated molecules, possibly due to its use of prior clusters
At the \emph{substructural} level, we report the percentage of molecules with certain types of rings defined by \citet{jiang2024pocketflow}, with respect to all ring-structured molecules. Pocket2Mol displays a preference for more fused rings as also noted by \citet{harris2023benchmarking}, while AR exhibits an obvious pattern in generating repeated three-membered rings. 

\begin{table}[htbp]
\caption{Percentage (\%) of molecular modes in terms of distribution and substructures. \emph{Note:} Fused refers to 80 specific rings,  3-Ring denotes three-membered rings, and so on. Highly deviated values are highlighted in \textbf{\textit{bold Italic}}.}
\label{tab:chemical_mode_collapse}
\resizebox{\linewidth}{!}{

\begin{tabular}{@{}lcccccc@{}}
\toprule
           & Unique            & Fused        & 3-Ring            & 4-Ring            & 5-Ring & 6-Ring \\ \midrule
Reference  & -                      & 30.0                     & 4.0                      & 0.0                      & 49.0          & 84.0          \\
Train      & -                      & 21.6                   & 3.8                    & 0.6                    & 56.1        & 90.9        \\
AR         & \textit{\textbf{36.2}} & 39.7 & \textit{\textbf{50.8}} & 0.8                    & 35.8        & 71.9        \\
Pocket2Mol & \textit{\textbf{73.7}} & \textit{\textbf{52.0}}   & 0.3                    & 0.1                    & 38.0          & 88.6        \\
FLAG       & 99.7                   & 42.4 & 3.1                    & 0.0                    & 39.9        & 84.7        \\
TargetDiff & 99.6                   & 37.8 & 0.0                      & \textit{\textbf{7.3}}  & 57.0          & 76.1        \\
Decomp-O   & \textit{\textbf{61.6}} & 13.1                   & 9.0                      & \textit{\textbf{11.4}} & 64.0          & 83.3        \\
Decomp-R   & \textit{\textbf{50.3}} & 28.1                   & 5.4                    & \textit{\textbf{8.3}}  & 51.5        & 65.6        \\
Ours       & 97.7                   & 30.9                   & 0.0                      & 0.6                    & 47.0          & 85.1        \\ \bottomrule
\end{tabular}

}
\end{table}

Geometrically measured, as shown in Fig.~\ref{fig:all_len}, atom-based autoregressive methods model the bond lengths for different bond types similarly, where reference distribution is multi-modal and varies across different types, while Pocket2Mol only captures a single mode, and for AR different bond lengths are distributed in a very similar fashion.


FLAG \citep{zhang_molecule_2023} generates fragment-by-fragment, which avoids collapsing by explicitly incorporating optimal and diverse substructures. But it suffers from more severe error accumulation, resulting in significant steric clashes and undesirable Vina Score (see Sec.~\ref{subsec:main}). 
Generally speaking, autoregressive models are still trapped in sub-optimal performance.
Intuitively, such limitations could be attributed to an unnatural atom ordering imposed during generation. 

\begin{figure}[htbp]
\begin{center}
\centerline{\includegraphics[width=\columnwidth]{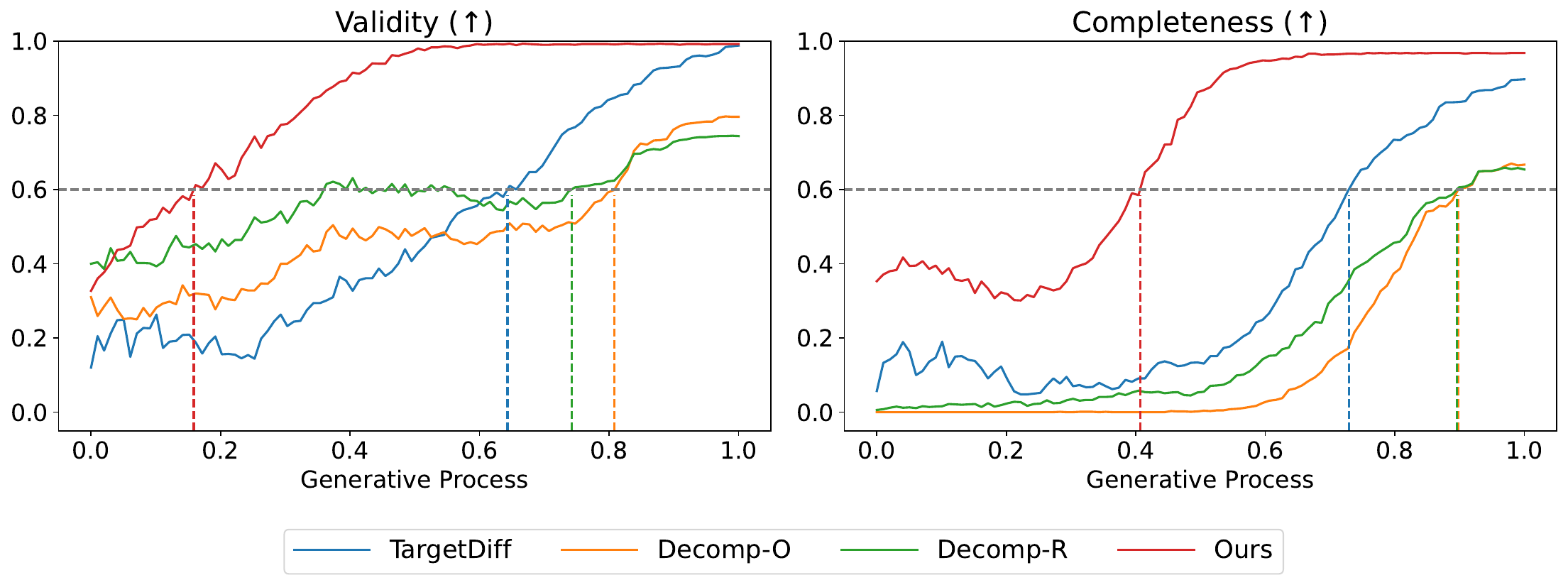}}
\caption{Percentage of valid, complete molecules in the trajectories during generative process.}
\label{fig:trajectory}
\end{center}
\vspace{-20pt}
\end{figure}

\subsection{Hybrid Continuous-Discrete Space}\label{subsec:hybrid_space}
Diffusion-based models, on the other hand, successfully alleviate mode collapse problem via non-autoregressive generation in terms of substructural distribution (see Fig.~\ref{fig:all_len}). However, the inconsistency between different modalities has long troubled molecular generation models, as suggested by MolDiff \citep{peng2023moldiff} and EquiFM~\citep{song2024equivariant}, where a careful design of either different noise levels or different probability paths is required.

A key insight is that the hybrid continuous-discrete space poses challenges to accurately capture the complicated data manifold for molecules, where the sample space in diffusion models is exposed to high variance, and the intermediate noisy latent is very likely to go outside the manifold.
Inspired by GeoBFN \citep{song2024unified}, we propose to operate within the fully continuous pamarater space, which enables considerably lower input variance and a smooth transformation towards the target distribution.

To further illustrate the difference between continuous-discrete diffusion and our fully continuous \ours, we sample 10 molecules for each of the 100 test proteins, and plot the curves of the ratio of valid, complete molecules against different timesteps during sampling. As shown in Fig.~\ref{fig:trajectory}, continuous-discrete diffusions heavily rely on the latter steps, passing a certain validity and completeness threshold in the final 60\%-90\% stage where noise scales are lower, while \ours~approaches target distribution far earlier (in the first 20\%-40\% steps), thereby possessing greater capacity to progressively refine and adjust the generated feasible structures, resulting in better conformations.

\section{Preliminary}
In this section, we briefly overview Bayesian Flow Networks (BFN) \citep{graves2023bayesian} in comparison with diffusion models for SBDD. For its detailed formulation and mathematical details, we refer readers to Appendix~\ref{sec:detailed_bfn}.

\begin{figure}[t]
\begin{center}
\centerline{\includegraphics[width=0.8\columnwidth]{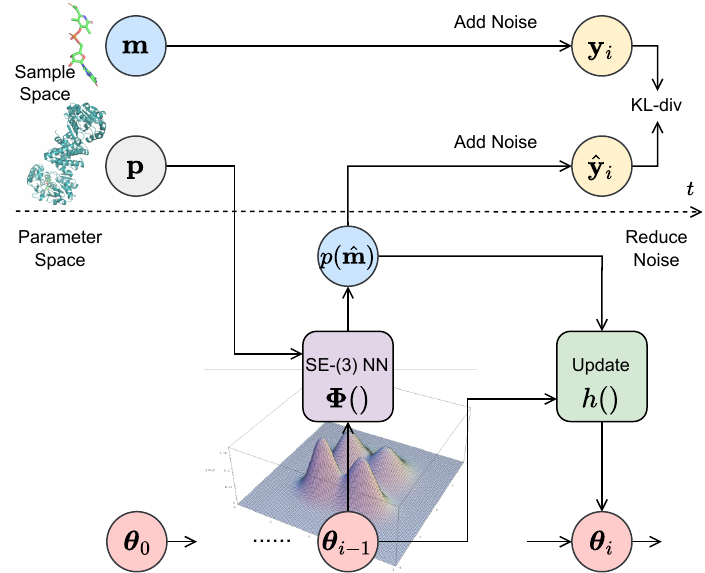}}
\caption{Overall Architecture.}
\label{fig:overall}
\end{center}
\end{figure}

\subsection{Problem Definition}
Structure-based Drug Design (SBDD) can be formulated as a conditional generation task. Given input protein binding site $\calP = \{(\bx_P^{(i)}, \bv_P^{(i)})\}_{i=1}^{N_P}$, which contains $N_P$ atoms with each $\bx_P^{(i)} \in \bbR^3$ and $\bv_P^{(i)} \in \bbR^{D_P}$ correspond to atom coordinates and atom features, respectively (\eg, element types, backbone or side chain indicator). 
The output is a ligand molecule $\calM = \{(\bx_M^{(i)}, \bv_M^{(i)})\}_{i=1}^{N_M}$, where $\bx_M^{(i)} \in \bbR^3$ and $\bv_M^{(i)} \in \bbR^{D_M}$, $N_M$ is the number of atoms in molecule. For convenience, we denote $\bp = [\bx_P, \bv_P], (\bx_P \in \bbR^{N_P \times 3}, \bv_P \in \bbR^{N_P \times D_P})$ and $\bm = [\bx_M, \bv_M], (\bx_M \in \bbR^{N_M \times 3}, \bv_M \in \bbR^{N_M \times D_M})$ as the concatenation of all protein or ligand atoms. 

\subsection{Molecular Generation in Parameter Space} 
The overall architecture of \ours~are shown in Fig.~\ref{fig:overall}. The generative process is viewed as message exchanges between a sender and a receiver, where the sender is only visible in sample space, and the receiver makes the guess from its understanding of samples and parameters. In every round of communication, the sender selects a molecule datapoint $\bm$, adds noise for timestep $t_i$ according to \textit{sender distribution} $p_S(\by_i~|~\bm; \alpha_i)$, and sends the noisy latent $\by$ to receiver, resembling the forward diffusion process. Here $\alpha_i$ is a noise factor from the schedule $\beta(t_i)$.

The receiver, on the other hand, outputs the reconstructed molecule $\hat\bm$ based on its previous knowledge of parameters $\btheta$, yielding \textit{output distribution} $p_O$. With the sender's noisy factor $\alpha$ known, the receiver can also add noise to the estimated output and give the predicted noisy latent, arriving at \textit{receiver distribution} $p_R$. 
\vspace{-20pt}

\begin{align}\label{eq:receiver}
p_R(\by_i~|~\btheta_{i-1}, \bp; t_i) = \underset{\hat\bm \sim p_O}{\bbE} p_S(\by_i~|~\hat\bm; \alpha_i), \\
\text{where} \quad p_O(\hat\bm~|~\btheta_{i-1}, \bp; t_i) = \bPhi(\btheta_{i-1}, \bp, t_i).\label{eq:p_o}
\end{align}
$\bPhi$ is a neural network which is expected to reconstruct clean sample $\hat\bm$ given parameters $\btheta_{i-1}$, pocket $\bp$ and time $t_i$.

The key difference between BFN and diffusion lies in its introduction of parameters. Thanks to structured Bayesian updates defined via Bayesian inference, the receiver is able to maintain fully continuous parameters and perform closed-form update on its belief of parameters. \textit{Bayesian update distribution} $p_U$ stems from the Bayesian update function $h$,
\vspace{-15pt}

\begin{equation}\label{eq:p_U}\footnotesize
p_U(\btheta_{i}~|~\btheta_{i-1}, \bm, \bp; \alpha_i) = \underset{\by_i' \sim p_S}{\bbE} \delta\Big(\btheta_i - h(\btheta_{i-1}, \by_{i}, \alpha_i)\Big),
\end{equation}
where $\delta(\cdot)$ is Dirac delta distribution. The parameter space enables arbitrarily applying noise as long as the Bayesian update is tractable, and eliminates the need to invert a pre-defined forward process as in diffusion models. 

According to the nice additive property of accuracy \citep{graves2023bayesian}, the \textit{Bayesian flow distribution} $p_F$ could be obtained to achieve simulation-free training, once teacher forcing with $\bm$ is applied:
\begin{align}\label{eq:p_f}
p_F(\btheta_i~|~\bm, \bp; t_i) & = \underset{\btheta_{1 \dots i-1} \sim p_U}{\bbE} p_U(\btheta_i~|~\btheta_{i-1}, \bm, \bp; \alpha_i) \nonumber \\
& = p_U(\btheta_i~|~\btheta_0, \bm, \bp; \beta(t_i))
\end{align}

Therefore, the training objective for $n$ steps is to minimize:
\begin{align}
L^n(\bm, \bp) = \underset{i \sim U(1, n)}{\bbE}~\underset{\by_i \sim p_S, \btheta_{i-1} \sim p_F}{\bbE}D_{\text{KL}}(p_S~\Vert~p_R).
\end{align}

\section{Methodology}
We introduce our proposed \ours~in as follows: in Sec.~\ref{subsec:param}, we demonstrate how to model continuous atom coordinates and discrete atom types within BFN framework, with the guarantee of SE-(3) equivariance for molecular data.
Then in Sec.~\ref{subsec:train_sample}, we elaborate our novel sampling strategy tailored for the parameter space. 
Within the fully continuous and differentiable space, \ours~is able to capture the global connection between different modalities, and sample efficiently with low variance.

\subsection{Resolving Different Modalities in Parameter Space}\label{subsec:param}
This section demonstrates how to resolve continuous atom coordinates and discrete atom types in parameter space.

\paragraph{Unified parameter $\btheta \defeq [\btheta^x, \btheta^v]$}
Following \citet{hoogeboom2022equivariant}, continuous atom coordinates $\bx$ are characterized by Gaussian distribution $\mathcal{N}(\bx~|~\bmu, \rho^{-1}\mathbf{I})$, and we set $\btheta^x = \left\{\bmu, \rho \right \}$, where $\bmu$ is learned and $\rho$ is pre-defined by noise factor $\alpha$. The Bayesian update function $\left\{\bmu_i, \rho_i \right \} \leftarrow h(\left\{\bmu_{i-1}, \rho_{i-1} \right \}, \by^x, \alpha_i)$ is defined as: 
\begin{align}
\rho_i &= \rho_{i-1} + \alpha_i \\
\bmu_i &= \frac{\bmu_{i-1}\rho_{i-1} + \by^x\alpha_i}{\rho_i}
\end{align}

For discrete atom types $v$, we use a categorical distribution $\btheta^v \in \bbR^{N_M \times K}$, and update it given $\alpha'$ via
\begin{align}
h(\btheta^v_{i-1}, \by^v, \alpha_i') & \defeq \frac{e^{\by^v}\btheta^v_{i-1}}{\sum_{k=1}^{K}e^{\by^v_k}(\btheta^v_{i-1})_k}
\end{align}
For prior $\btheta_0$, we adopt standard Gaussian and uniform distribution respectively, following \citet{graves2023bayesian}. 


\paragraph{Applying noise for different modalities}
Thanks to the continuous nature of parameters, we are able to apply the following continuous noise even for discrete atom types,
instantiating the \textit{sender distribution} $p_S$:
\begin{align}
    p_S(\by^x~|~\bx_M; \alpha) &= \mathcal{N}(\by^x~|~\bx_M, \alpha^{-1}\mathbf{I}) \\
    p_S(\by^v~|~\bv_M; \alpha') &= \mathcal{N}\Big(\by^v~|~\alpha'(K \be_{\bv_M} - \mathbf{1}), \alpha' K \mathbf{I}\Big)
\end{align}
where $\be_{\bv_M} = [\be_{\bv_M^{(1)}}, \dots, \be_{\bv_M^{(N_M)}}] \in \bbR^{N_M \times K}$, $\be_j \in \bbR^{K}$ is the projection from the class index $j$ to the length-$K$ one-hot vector, and $K$ the number of atom types.
Note that we could set different noise schedules for different modalities ($\alpha$ for coordinates and $\alpha'$ for types) for more efficient training of the joint noise prediction network.

Thereby for \textit{receiver distribution} in Eq.~\ref{eq:receiver},
\begin{align}\footnotesize
& p_R(\by^x~|~\btheta^x, \bp; t) = \mathcal{N}(\by^x~|~\bPhi(\btheta^x, \bp, t), \alpha^{-1}\mathbf{I}) \\
& p_R(\by^v~|~\btheta^v, \bp; t) = \Big[p_R\big((\by^v)^{(d)}|\cdot\big)\Big]_{d = 1 \dots N},
\end{align}
where $
p_R\Big((\by^v)^{(d)}|\cdot\Big) = \sum_{k} p_O^v(k|\cdot) p_S^v\Big((\by^v)^{(d)} | k; \alpha\Big).
$


\paragraph{SE-(3) equivariance}
We introduce a fundamental inductive bias for SBDD to BFN, \ie~the density should be invariant to translation and rotation of protein-molecule complex \cite{satorras2021n, xu2021geodiff, hoogeboom2022equivariant}, in the following proposition (proof in Appendix~\ref{sec:se3_inv}). 
\begin{proposition}
Denote the SE-(3) transformation as $T_g$, the likelihood is invariant \emph{w.r.t.} $T_g$ on the protein-ligand complex: $p_\phi(T_g(\bm | \bp)) = p_\phi(\bm | \bp)$ if we shift the Center of Mass (CoM) of protein atoms to zero and parameterize $\bPhi(\btheta, \bp, t)$ with an SE-(3) equivariant network.
\end{proposition}

\subsection{Noise Reduced Sampling in Parameter Space}\label{subsec:train_sample}
\ours~addresses the high-variance discrete variable problem by maintaining a continuous probability mass function as beliefs of distributional parameters, which allows a smooth transformation towards the target distribution.
This natural coherence with continuous coordinates gives us an advantage over continuous-discrete diffusion process.

During sampling, original BFN shifts the denoising process from sample space (recall diffusion $\by_{i-1} \rightarrow \by_i$) to parameter space $(\btheta_{i-1}, \by_i) \rightarrow \btheta_i$ via Bayesian update function $h$, where the information flows in this direction:
\begin{align}
{\btheta_{i-1}} \xrightarrow[]{\bPhi} \hat\bm \xrightarrow[]{p_S} \by_i \xrightarrow[]{p_U} \btheta_i,
\end{align}
where $p_U(\btheta_{i}~|~\btheta_{i-1}, \bm, \bp; \alpha_i)$ is defined in Eq.~\ref{eq:p_U}, and $\bm$ is set to estimated $\hat\bm$ drawn from $p_O$ in Eq.~\ref{eq:p_o}.

It should be noted that the existing generative process of BFN, as well as that of diffusion models, performs continuous atom coordinates and discrete atom type sampling at each timestep. This risks introducing too much noise, and might end up generating incomplete molecules. To alleviate such a problem, we design an empirically effective sampling strategy, which operates within the parameter space, and thus avoids introducing further noise from sampling discrete variables. The graphical description becomes:

\begin{align}
\btheta_{i-1} \xrightarrow[]{\bPhi} \hat\bm \xrightarrow[]{p_F} \btheta_i
\end{align}

Specifically, denoting $\gamma(t) \defeq \frac{\beta(t)}{1-\beta(t)}$, we update the parameter via Eq.~\ref{eq:p_f}, which simplifies to:
\begin{align}\footnotesize
    & p_F(\bmu~|~\hat\bx, \bp; t) = \mathcal{N}\Big(\bmu~|~\gamma(t)\hat\bx, \gamma(t)(1-\gamma(t))\bold{I}\Big) \\
    & \quad p_F(\btheta^v~|~\hat\bv, \bp; t) \nonumber \\
    & = \underset{\mathcal N\big(\by^v | \beta(t)(K\be_{\hat\bv} - \mathbf{1}), \beta(t)K\mathbf{I}\big)}{\bbE} \delta(\btheta^v - \text{softmax}(\by^v))
\end{align}
We use the estimated $\hat\bm = [\hat\bx, \hat\bv]$ (note that $\hat\bv$ directly takes the continuous output categorical values without sampling) to directly update parameter for the next step, bypassing the sampling of noisy data needed for Bayesian update $\btheta_i = h(\btheta_{i-1}, \by, \alpha)$. The whole generative process happens in the parameter space except for the final step, which enjoys the advantage of lower variance and accelerates the overall generation path towards the complicated structure of molecules, with greatly improved sample quality at significantly fewer sampling steps, as shown in Fig.~\ref{fig:ablation}. Details of sampling are described in Algorithm~\ref{algo:sample}.

\section{Experiments}

\begin{table*}[t]
\caption{Summary of different properties of reference and generated molecules under different sizes. (↑) / (↓) indicates larger / smaller is better. Top 2 results are highlighted with \textbf{bold text} and \ul{underlined text}. \emph{Note:} SE is short for Strain Energy, Div for Diversity, BF for Binding Feasibility, and SR for Success Rate. Baselines are either evaluated based on publicly available codebase (Decomp-R) or officially released samples (others).}
\label{tab:main_results}
\begin{center}
\resizebox{\linewidth}{!}{
\begin{tabular}{@{}l|cccccc|ccccc|ccc|c|cc@{}}
\hline
\multirow{4}{*}{Methods} & \multicolumn{6}{c|}{\multirow{2}{*}{\color{black}{Binding Affinity}}}                                                                                                      & \multicolumn{5}{c|}{\color{black}{Conformation Stability}}                                                                        & \multicolumn{3}{c|}{\multirow{2}{*}{\color{black}{Drug-like Properties}}}                                        & \multicolumn{2}{c|}{{\multirow{2}{*}{\color{black}{Overall}}}} &                      \\ \cline{8-12} 
                         & \multicolumn{6}{c|}{}                                                                                                                                       & \multicolumn{3}{c|}{Ligand}                & \multicolumn{2}{c|}{Complex}                                     & \multicolumn{3}{c|}{}                                         & \multicolumn{2}{c|}{} 
                         & \multicolumn{1}{l}{} \\ \cline{2-18} 
                         & \multicolumn{2}{c|}{Vina Score ($\downarrow$)}       & \multicolumn{2}{c|}{Vina Min ($\downarrow$)}         & \multicolumn{2}{c|}{Vina Dock ($\downarrow$)} & \multicolumn{3}{c|}{SE ($\downarrow$)}          & \multicolumn{1}{c|}{Clash ($\downarrow$)} & RMSD ($\uparrow$)             & \multicolumn{1}{c|}{SA ($\uparrow$)} & \multicolumn{1}{c|}{QED ($\uparrow$)} & Div ($\uparrow$) & {BF ($\uparrow$)}  & \multicolumn{1}{c|}{SR ($\uparrow$)}                & Size                 \\
                         & Avg.           & \multicolumn{1}{c|}{Med.}           & Avg.           & \multicolumn{1}{c|}{Med.}           & Avg.                  & Med.                  & 25\%        & 50\% & \multicolumn{1}{c|}{75\%}         & \multicolumn{1}{c|}{Avg.}                 & $\%<$ 2 \r{A} & \multicolumn{1}{c|}{Avg.}            & \multicolumn{1}{c|}{Avg.}             & Avg.             & ($\%$) & \multicolumn{1}{c|}{($\%$)}            & Avg.                 \\ \hline 
Reference                & -6.36          & \multicolumn{1}{c|}{-6.46}          & -6.71          & \multicolumn{1}{c|}{-6.49}          & -7.45                 & -7.26                 & 34    & 107      & \multicolumn{1}{c|}{196}          & \multicolumn{1}{c|}{5.51}                 & 34.0                 & \multicolumn{1}{c|}{0.73}            & \multicolumn{1}{c|}{0.48}             & -                & 29.0   & \multicolumn{1}{c|}{25.0}                  & 22.8                 \\ \hline 
AR                       & \ul{-5.75}     & \multicolumn{1}{c|}{\ul{-5.64}}          & -6.18          & \multicolumn{1}{c|}{\ul{-5.88}}          & -6.75                 & -6.62                 & 259 & 595         & \multicolumn{1}{c|}{2286}         & \multicolumn{1}{c|}{\textbf{4.49}}        & \ul{31.1}                 & \multicolumn{1}{c|}{0.64}            & \multicolumn{1}{c|}{0.51}             & \ul{0.70}             & 17.3  & \multicolumn{1}{c|}{7.1}                   & 17.7                 \\
Pocket2Mol               & -5.14          & \multicolumn{1}{c|}{-4.70}          & \textbf{-6.42}          & \multicolumn{1}{c|}{-5.82}          & \textbf{-7.15}                 & \textbf{-6.79}                 & \ul{102} & \ul{189} &  \multicolumn{1}{c|}{\ul{374}} & \multicolumn{1}{c|}{6.24}            & 30.8                 & \multicolumn{1}{c|}{\textbf{0.76}}   & \multicolumn{1}{c|}{\textbf{0.57}}        & 0.69             & \ul{24.6}       & \multicolumn{1}{c|}{\textbf{24.4}}         & 17.7                 \\
Ours-small             & \textbf{-5.96} & \multicolumn{1}{c|}{\textbf{-5.89}} & \ul{-6.33}    & \multicolumn{1}{c|}{\textbf{-6.04}} & \ul{-6.98}    & \multicolumn{1}{c|}{\ul{-6.63}}    & \textbf{44}   & \textbf{103}     & \multicolumn{1}{c|}{\textbf{274}}       & \multicolumn{1}{c|}{\ul{4.88}}    & \multicolumn{1}{c|}{\textbf{38.6}}         & \multicolumn{1}{c|}{\ul{0.74}}    & \multicolumn{1}{c|}{\ul{0.52}}    & \textbf{0.74} & \textbf{33.3}      &  \multicolumn{1}{c|}{\ul{17.4}}     & 17.8          \\
\hline 
FLAG\tablefootnote{Though relatively worse than reported in FLAG paper, the evaluation is based on released samples and confirmed by authors.}                     & 16.48 & \multicolumn{1}{c|}{4.53} & 1.21 & \multicolumn{1}{c|}{-4.04} & -5.63 & -6.61 & 143 & \ul{396} & \multicolumn{1}{c|}{\ul{1164}} & \multicolumn{1}{c|}{40.83}                & 8.2                  & \multicolumn{1}{c|}{\textbf{0.70}}            & \multicolumn{1}{c|}{{0.49}}    & 0.70             & 3.8        & \multicolumn{1}{c|}{14.1}              & 21.5                 \\ 
TargetDiff               & \ul{-5.47}          & \multicolumn{1}{c|}{\ul{-6.30}}     & \ul{-6.64}          & \multicolumn{1}{c|}{\ul{-6.83}}          & \ul{-7.80}            & \ul{-7.91}                 & 369 & 1243        & \multicolumn{1}{c|}{13871}        & \multicolumn{1}{c|}{10.84}                & \ul{29.4}            & \multicolumn{1}{c|}{0.58}            & \multicolumn{1}{c|}{0.48}             & \ul{0.72}        & 14.4              & \multicolumn{1}{c|}{10.5}       & 24.2                 \\
Decomp-R                 & -5.19          & \multicolumn{1}{c|}{-5.27}          & -6.03          & \multicolumn{1}{c|}{-6.00}          & -7.03                 & -7.16                 & \ul{115}  & 421         & \multicolumn{1}{c|}{{1424}}         & \multicolumn{1}{c|}{\ul{8.16}}                 & 22.7                 & \multicolumn{1}{c|}{{0.66}}            & \multicolumn{1}{c|}{\textbf{0.51}}             & \textbf{0.73}    & \ul{14.6}            & \multicolumn{1}{c|}{\ul{14.9}}         & 21.2                 \\
Ours                     & \textbf{-6.59} & \multicolumn{1}{c|}{\textbf{-7.04}} & \textbf{-7.27} & \multicolumn{1}{c|}{\textbf{-7.26}} & \textbf{-7.92}            & \textbf{-8.01}            & \textbf{83} &  \textbf{195}  & \multicolumn{1}{c|}{\textbf{510}}     & \multicolumn{1}{c|}{\textbf{7.09}}                 & \textbf{41.8}        & \multicolumn{1}{c|}{\ul{0.69}}       & \multicolumn{1}{c|}{\ul{0.50}}             & \ul{0.72}        & \textbf{33.8}     & \multicolumn{1}{c|}{\textbf{26.8}}       & 22.7                 \\ \hline
Decomp-O                 & -5.67          & \multicolumn{1}{c|}{-6.04}          & -7.04     & \multicolumn{1}{c|}{-7.09}     & -8.39        & -8.43        & 379 & 983         & \multicolumn{1}{c|}{4133}         & \multicolumn{1}{c|}{14.63}                & 23.9                 & \multicolumn{1}{c|}{0.61}            & \multicolumn{1}{c|}{0.45}             & \textbf{0.68}             & 11.1         & \multicolumn{1}{c|}{24.5}            & 29.4                 \\
Ours-large             & \textbf{-6.61} & \multicolumn{1}{c|}{\textbf{-8.14}} & \textbf{-8.14} & \multicolumn{1}{c|}{\textbf{-8.42}} & \textbf{-9.25} & \multicolumn{1}{c|}{\textbf{-9.20}} & \textbf{171} & \textbf{333} &     \multicolumn{1}{c|}{\textbf{1110}}      & \multicolumn{1}{c|}{\textbf{10.73}} & \multicolumn{1}{c|}{\textbf{42.7}}         & \multicolumn{1}{c|}{\textbf{0.62}} & \multicolumn{1}{c|}{\textbf{0.46}} & 0.61          & \textbf{31.1}      & \multicolumn{1}{c|}{\textbf{36.1}}      & 29.4         \\ \hline
\end{tabular}

}
\end{center}
\vspace{-10pt}
\end{table*}

\subsection{Experimental Setup}

\paragraph{Dataset} 
We use the CrossDocked dataset \citep{doi:10.1021/acs.jcim.0c00411} for training and testing, which originally contains 22.5 million protein-ligand pairs, and after the RMSD-based filtering and 30\% sequence identity split by \citet{luo_3d_2022}, results in 100,000 training pairs and 100 test proteins. For each test protein, we sample 100 molecules for evaluation.

\paragraph{Baselines} 
For autoregressive sampling-based models, we choose atom-based models
AR \citep{luo_3d_2022}, Pocket2Mol \citep{peng_pocket2mol_2022} and fragment-based model FLAG
\citep{zhang_molecule_2023}. 
For diffusion-based models, we consider TargetDiff \citep{guan_3d_2023} and two variants of DecompDiff \citep{guan_decompdiff_2023}. Decomp-R uses the prior estimated from reference molecules in the test set, while Decomp-O selects the optimal prior from the reference prior and pocket prior, where the pocket prior center is predicted by AlphaSpace2 \cite{doi:10.1021/acs.jcim.9b00652} and ligand atom number by a neural classifier.





\paragraph{Evaluation} 
We conduct a comprehensive evaluation of SBDD models on all 100 proteins in test set, including:
\begin{itemize}
\item \textbf{\color{black}{Binding Affinity}}. We employ AutoDock Vina \citep{https://doi.org/10.1002/jcc.21334} to measure binding affinity as it is a common practice \citep{luo_3d_2022, peng_pocket2mol_2022, guan_3d_2023, guan_decompdiff_2023}, and report \textbf{Vina Score}, a direct score of generated pose, \textbf{Vina Min}, which scores the optimized pose after a local minimization of energy, and \textbf{Vina Dock}, the best possible score after re-docking, a global grid-based search optimization process. Therefore, it is highly favorable if Vina Score is close to Vina Min and Vina Dock, suggesting that the generated poses capture the 3D interaction well.
\item \textbf{\color{black}{Conformation Stability}}. We measure the stability for \emph{ligand-only} and \emph{binding complex} conformation. For \emph{ligand-only}, we use the Jensen-Shannon divergence (\textbf{JSD}) between reference and generated distributions of bond length, bond angle and torsion angle at substructure level, and for a more global view, we employ \textbf{Strain Energy} to evaluate the rationality of generated ligand conformation. For \emph{binding complex}, we adopt Steric Clashes (\textbf{Clash}) to detect possible clashes in protein-ligand complex, following \citet{harris2023benchmarking}. We further propose to evaluate symmetry-corrected \textbf{RMSD} between the generated ligand atoms and Vina redocked poses as the metric of binding mode consistency, where poses with an RMSD below 2\r{A} is generally regarded as chemically meaningful \cite{alhossary2015fast, hassan2017protein, mcnutt2021gnina}.
\item \textbf{\color{black}{Drug-like Properties}}. Drug-likeliness (\textbf{QED}), synthetic accessibility (\textbf{SA}), and diversity (\textbf{Div}) are adopted as molecular property metrics.
\item \textbf{\color{black}{Overall}}. To evaluate the overall quality of generated molecules, we calculate the \textbf{Binding Feasibility} as the ratio of molecules with reasonable affinity (Vina Score $<$ -2.49 kcal/mol) 
and stable conformation (Strain Energy $<$ 836 kcal/mol, RMSD $<$ 2\r{A}) simultaneously, where the threshold values are set to the 95 percentile of the reference molecules. We also report \textbf{Success Rate} (Vina Dock $<$ -8.18, QED $>$ 0.25, SA $>$ 0.59) following \citet{long2022DESERT} and \citet{guan_3d_2023}.
\item \textbf{\color{black}{Sample Efficiency}}. In order to make a practical comparison among non-autoregressive methods, we report the average \textbf{Time} and \textbf{Generation Success}, with the latter defined as the ratio of valid and complete molecules versus the intended number of samples. 
\end{itemize}

\subsection{Main Results}\label{subsec:main}

Our main findings are listed as below:
\begin{itemize}
    \item \ours~resembles and even surpasses the reference set in terms of binding affinity and overall feasibility, showing that we effectively learn the binding dynamics from protein-ligand complex distribution.
    \item Non-autoregressive molecule generation could benefit from modeling in continuous parameter space, demonstrated by our performance in capturing diverse substructural modes and greatly improved conformation.
    \item Reliable evaluation of SBDD ought to take molecule sizes into account. To achieve fair comparison, controlled experiment regarding molecule size is needed.
\end{itemize}

\paragraph{\color{black}{Binding Affinity}}
We report Vina metrics in Table~\ref{tab:main_results}. 

\emph{I. Our model consistently outperforms other strong baselines in affinities}, achieving a reference-level Vina Score of -6.59 kcal/mol. As Vina Score directly scores the pose and Vina Min only optimizes locally, they directly measure the generated pose quality. To the best of our knowledge, \ours~is the first to achieve reference-level affinity scores without significant rearrangements via redocking, showing our superiority in learning binding interactions.


\emph{II. Vina Dock can potentially be hacked by generating larger molecules.} Intuitively, larger molecules have more chances of forming interactions with protein surfaces.
With the largest molecule sizes, Decomp-O achieves the second-best Vina Dock (-8.39 kcal/mol), far better than reference molecules. Further investigation reveals that Decomp-O gains an advantage by producing considerably larger out-of-distribution (OOD) molecules and thereby brings up the highest possible affinity post-docking. For a fair comparison, we report variants of DecompDiff and \ours~stratified by size, and with the same number of atoms as Decomp-O, our model consistently achieves SOTA affinities, underscoring its robustness across different molecular sizes. 

\begin{table}[tbp]
\caption{Summary of molecular conformation results. (↓) indicates smaller is better. Top 2 results are highlighted with \textbf{bold text} and \ul{underlined text}. \emph{Note:} JSD is calculated between distributions estimated from generated and reference molecules, we report the mean of all JSD values here.}
\label{tab:posecheck_results}
\begin{center}
\scriptsize

\begin{tabular}{l|c|c|c}
\toprule
\multirow{2}{*}{Methods} & Length ($\downarrow$) & Angle ($\downarrow$) & Torsion ($\downarrow$)\\
                         & Avg. JSD                    & Avg. JSD                   & Avg. JSD                                  \\ \midrule
AR                       & 0.554                    & 0.507                   & 0.552                            \\
Pocket2Mol               & 0.485                    & 0.482                   & 0.459                                \\
FLAG                     & 0.511                    & \ul{0.406}             & \textbf{0.270}                             \\ \midrule
TargetDiff               & 0.382                    & 0.435                   & 0.400                                    \\
Decomp-O                 & 0.359                    & 0.414                   & 0.358                                    \\
Decomp-R                 & \ul{0.348}              & 0.412                   & 0.317                                       \\
Ours                     & \textbf{0.319}           & \textbf{0.379}          & \ul{0.300}                               \\  \bottomrule
\end{tabular}
\end{center}
\end{table}

\paragraph{\color{black}{Conformation Stability}} 
We report the \emph{substructural} level's average Jensen-Shannon divergence (JSD) between reference and generated bond length, angle and torsion angle distributions in Table~\ref{tab:posecheck_results} (detailed results for different bond/angle/torsion types in Appendix~\ref{sec:exp_detail}). 
At the \emph{global structure} level, we report strain energy for ligand-only conformational stability, and measure clashes in the binding complex, together with RMSD between generated and redocked poses in Table~\ref{tab:main_results}.

\emph{I. Our model excels in modeling diverse local modes}, and ranks first in bond length and angle distributions. 
Moreover, Fig.~\ref{fig:all_len} shows \ours~is the only model that captures two distinct modes for multi-modal C-C, C-N and C-O bond, justifying our choice of modeling in the joint continuous parameter space. More results are in Fig.~\ref{fig:other_len}, \ref{fig:other_angle} and \ref{fig:other_torsion}.

\emph{II. Injecting substructural inductive bias helps to capture more modes.} Fragment-based model FLAG displays the best torsion angle distribution, and prior-enhanced DecompDiff also exhibits relatively competitive performances in modeling molecular geometries, whereas other autoregressive models collapse into certain modes as in Fig.~\ref{fig:all_len}. 

\emph{III. For ligand-only stability, we greatly improve upon the strained conformations}, even surpassing autoregressive methods.
According to Table~\ref{tab:main_results}, our model is at least an order of magnitude better than diffusion-based counterparts, and is close to reference. 
While autoregressive methods generally display better strain energy, \ours~still achieves superior performance under comparable molecule sizes.

\emph{IV. Our binding complex contains fewer clashes and remains consistent after redocking.}
We achieve few steric clashes, and has the best RMSD performance, which means 46\% of our molecules already resemble accurate docking pose even without force field optimization or redocking, rendering it reliable for generating molecules in 3D space. 
The reason why we achieve even better RMSD than reference can be explained by a distribution shift. In dataset construction\footnote{There are two kinds of 3D ligand poses in the dataset, i.e. Vina minimized poses in the given receptor, and Vina docked poses. \url{https://github.com/gnina/models/tree/master/data/CrossDocked2020}}, the training set contains 52.4\% docked molecules, while the test set only contains 37.0\% docked ones, which aligns with the fact that for reference there are only 34.0\% with RMSD $<$ 2\r{A}. This accounts for why \ours~has more consistent and high-affinity binders, which effectively captures the training set distribution and learns the binding dynamics.

\paragraph{\color{black}{Overall}}
We report the overall feasible rate and success rate in Table~\ref{tab:main_results}.
\ours~achieves the best among all, demonstrating our competency in generating molecules with
high affinity and stable conformation. Our method captures the interatomic interactions in 3D space, and proposes desirable molecules without relying on post-fixed docking poses. This further validates our choice of learning in the continuous parameter space.





\begin{figure}[tbp]

    \centering
    \includegraphics[width=0.8\linewidth]{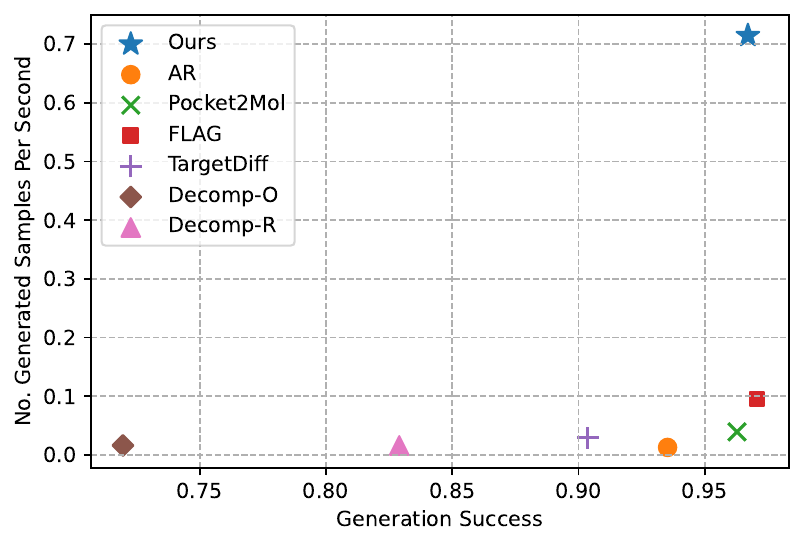}
    \caption{Sample efficiency, where Generation Success means the generated molecules are both valid and complete.}
    \label{fig:sample_efficiency}
    
\end{figure}

\paragraph{\color{black}{Sampling performance}}
We compare the generation speed (average time for generating 100 samples) and generation success in Figure~\ref{fig:sample_efficiency}. We achieve SOTA sampling performance in both dimensions, generating more complete (96.7\%) molecules at \textbf{$30 \times$ speedup}. While it takes on average 3428s and 6189s for TargetDiff and DecompDiff to generate 100 samples respectively, our model only uses 141s, thanks to our improved sampling strategy (see Sec.~\ref{subsec:ablation}). 

\subsection{Ablation Study of Sampling Strategy}\label{subsec:ablation}

Considering that we propose the first-of-its-kind SBDD model that operates in the fully continuous parameter space, and present a noise-reduced sampling approach adapted to the space, we conduct ablation study that validates our design, showing a performance boost from Vina Score/Min of -5.42/-6.30 kcal/mol to -6.51/-7.13 kcal/mol.

We test different sampling strategies with different steps for the same checkpoint, and sample 10 molecules each for 100 test proteins. We plot the curves of QED, SA, Completeness $(\uparrow)$ and Vina Score $(\downarrow)$ in Figure~\ref{fig:ablation}, Appendix~\ref{subsec:app_ablation}.
As the sampling step increases to training steps, we found the original sampling strategy exhibits first enhanced then slightly decreased sample quality, possibly because the update of parameters is smoothed or oversmoothed by finer partitioned noise factor $\alpha$, whereas the noise reduced strategy displays this tendency far earlier and generates the best quality of molecules with fewer sampling steps, indicating its high efficiency. Considering the overall sample quality, we decide to use 100 sampling steps for our model, which is $10\times$ faster than sampling at original 1000 training steps.





\section{Related Work}

\paragraph{Target-Aware Molecule Generation} 
Trained on protein-ligand complex data, target-aware methods directly model the interaction between protein pockets and ligands. Early attempts are based on 1D SMILES or 2D molecular graph generation \cite{bjerrum2017molecular, gomez2018automatic, segler2018generating} and fail to consider spatial information. Recent works focus on 3D molecule generation, and there are mainly two fashions: 
(1) \emph{Autoregressive methods}. 
For atom-based methods, LiGAN \citep{masuda_generating_2020} and AR \citep{luo_3d_2022} adopt an atomic density grid view of molecules, the former predicting a voxelized density grid and performing optimization to reconstruct atom types and coordinates, the latter assigning atomic probability to each voxel and utilizes MCMC to generate atom-by-atom. 
GraphBP \citep{liu_generating_2022} uses normalizing flow and encodes the context to preserve 3D geometric equivariance, and Pocket2Mol \citep{peng_pocket2mol_2022} further adds bond generation for more realistic molecular structure. 
For fragment-based methods \citep{powers2022fragmentbased, zhang_learning_2023, zhang_molecule_2023}, molecules are decomposed into chemically meaningful motifs rather than seperated atom point cloud, and generated via motif assembling. 
(2) \emph{Diffusion-based methods} have recently been proposed, aiming to overcome the problem of sampling efficiency and unnatural ordering brought by autoregressive fashion \citep{schneuing_structure-based_2022, guan_3d_2023, guan_decompdiff_2023}.
But these methods still suffer from false positive problems.

\section{Conclusion}
In this paper, we first investigate the challenges of current generative models in SBDD, \ie, distorted structures and sub-optimal binding modes. 
Based on the observations concerning mode collapse and hybrid space, we propose \ours, an SE-(3) equivariant generative model operating in the continuous parameter space with a noise reduced sampling strategy, 
which yields higher quality molecules.

\section*{Acknowledgements}
This work is supported by the National Science and Technology Major Project (2022ZD0117502), Natural Science Foundation of China (62376133) and Guoqiang Research Institute General Project, Tsinghua University (No. 2021GQG1012). 
The authors would like to thank Xiangyu Li for his valuable comments on this work.

\section*{Impact Statement}
This paper is aimed to facilitate in-silico rational drug design. Potential society consequences include mal-intended usage of toxic compound discovery, which needs support from professional wet labs and thus expensive to reach. Therefore we do not possess a negative vision that this might lead to serious ethical consequences, though we are aware of such a possibility.


\bibliography{example_paper}
\bibliographystyle{icml2024}

\newpage
\appendix
\onecolumn
\section{Detailed Formulation of BFN}\label{sec:detailed_bfn}

\paragraph{Introducing parameters to diffusion process} 
Classic diffusion process consists of a \textit{forward process} which gradually applies noise to the data $\by_i \sim p(\by_i | \bm; \alpha_i)$ till finally standard Gaussian $\by_0 \sim \mathcal N(\mathbf{0}, \mathbf{I})$, and a \textit{reverse process} which starts from Gaussian noise $\by_0$ and iteratively denoises $\by_{i} \sim p(\by_i | \by_{i-1}, \bp; t)$ to produce a sample $\bm \sim p(\bm|\by_n, \bp; n)$. Thus the key designs of diffusion are about how to apply noise given $\bm$, and how to denoise with $(\by, \bp)$.

The \textit{Variational Lower Bound} for diffusion models \cite{ho2020denoising}
is:
\begin{align}\label{eq:objective}
- \log p_\theta(\bm|\bp) \le  \mathcal L_{\text{VLB}} = D_{\text{KL}}\Big(q(\by_{0 \dots n}|\bm, \bp) \Vert p_\phi(\by_{0 \dots n}|\bp)\Big) 
\end{align}
Through adding noise to $\bm$ and denoising from $\by_0$ and $\bp$, diffusion transports a simple prior distribution $p(\by_0)$ to the desired data distribution $p(\bm | \bp)$, and generates target-aware molecules in a non-autoregressive fashion.
To transfer the generative process from sample space to parameter space, latent variables $\btheta_{0 \dots n}$ are further introduced to characterize the distribution of $\by_{1 \dots n}$ \citep{graves2023bayesian}:
\begin{align}\label{eq:step_kl}\footnotesize
\mathcal L_{\text{VLB}} = D_{\text{KL}}\Big(q(\by_{1 \dots n}, \btheta_{0 \dots n}|\bm, \bp) \Vert p_\phi(\by_{1 \dots n}, \btheta_{0 \dots n}|\bp)\Big) = n \underset{i \sim U(1, n)}{\bbE} \underset{\by_i, \btheta_{i-1} \sim q}{\bbE} D_{\text{KL}} \Big(q(\by_i|\bm) \Vert p_\phi(\by_i | \btheta_{i-1}, \bp)\Big) .
\end{align}
With Eq.~\ref{eq:step_kl} pulling close $p_\phi$ to $q$, $p_\phi$ is finally able to alternatively generates $\by_i$ and $\btheta_i$, \ie, a generative process driven from the parameter space. Here for BFN, the generative process is characterized by  \textit{receiver distribution} (Eq.~\ref{eq:receiver}) with the help of \textit{output distribution} (Eq.~\ref{eq:p_o}), and the noise adding process by \textit{sender distribution}:

\begin{align}\label{eq:sender}
q(\by_i~|~\bm) = p_S(\by_i~|~\bm; \alpha_i)
\end{align}

\paragraph{Training objective}
BFN can be trained by minimizing the KL-divergence between noisy sample distributions.
BFN 
allows training in discrete time and continuous time, and for efficiency we adopt the $n$-step discrete loss. Since the atom coordinates and the noise are Gaussian, the loss can be written analytically as follows:
\begin{align}\footnotesize
    L^n_x &= D_{\mathrm{KL}}\Big(\mathcal{N}(\bx,\alpha_i^{-1}I) ~\Vert~ \mathcal{N}(\hat{\bx}(\btheta_{i-1},\bp,t),\alpha_i^{-1}I)\Big) \nonumber \\
    &= \frac{\alpha_i}{2}\Big\Vert \bx-\hat{\bx}(\btheta_{i-1},\bp,t)\Big\Vert^2
\end{align}
Similarly, atom type loss can also be derived by taking KL-divergence between Gaussians, yielding:
\begin{align}\footnotesize
    L^n_v &= \ln\mathcal{N}\Big(\by^v~|~\alpha_i(K\be_{\bv}-\mathbf{1}),\alpha_iKI\Big)- \\
&\sum_{d=1}^{N_M}\ln\Big(\sum_{k=1}^{K}p_O(k|\btheta;t)\mathcal{N}\big(\cdot^{(d)}|\alpha_i(K\be_k-\mathbf{1}),\alpha_iKI\big)\Big) \nonumber
\end{align}


\begin{algorithm}[H]
\caption{Discrete-Time Loss}\label{algo:dloss}
\begin{algorithmic}[1]
\REQUIRE $\bx_M \in \bbR^{3N_M}, \bv_M \in \bbR^{N_MK}, \bp \in \bbR^{N_P(3+D_P)}, \sigma_1, \beta_1 \in \bbR^+$
\STATE $i \sim U(1, n), t \gets \frac{i-1}{n}$ 
\STATE $\bmu \sim p_F^x(\bmu|\bx_M, \bp; t, \beta(t) = \sigma_1^{-2t} - 1)$
\STATE $\btheta^v \sim p_F^v(\btheta^v | \bv_M, \bp; t, \beta(t) = t^2 \beta_1)$
\STATE $\hat{\bx}, \hat\bv \gets p_O(\bmu, \btheta^v, \bp, t)$
\STATE $L^n_x \gets \frac{(1 - \sigma_1^{2/n})}{2\sigma_1^{2i/n}} \Vert \bx_M - \hat{\bx}\Vert^2$
\STATE $\alpha \leftarrow \beta_1(\frac{2i-1}{n^2})$
\STATE $\by^v \sim p_S^v(\by^v | \bv_M, \alpha))$
\STATE $L^n_v \gets \ln p_S^v(\by^v|\bv_M; \alpha) - \ln p_R^v(\by^v|\hat\bv; \alpha, t)$
\STATE return $L^n(\bm, \bp) = L^n_x + L^n_v$
\end{algorithmic}
\end{algorithm}

\begin{algorithm}[H]
\caption{Sampling}\label{algo:sample}
\begin{algorithmic}[1]
\FUNCTION{$\text{update}({\hat\bx} \in \bbR^{3N}, \hat\bv \in \bbR^{NK}, \beta(t), \beta'(t), t \in \bbR^+)$} 
\STATE {$\gamma \gets \frac{\beta(t)}{1 - \beta(t)}$}
\STATE {$\bmu \sim \mathcal N(\gamma\hat\bx, \gamma(1-\gamma)\mathbf{I})$}
\STATE {$\by^v \sim \mathcal N(\by^v | \beta'(t)(K\be_{\hat\bv} - \mathbf{1}), \beta'(t)K\mathbf{I})$}
\STATE {$\btheta^v \gets [\text{softmax}((\by^v)^{(d)})]_{d = 1 \dots N_M}$}
\STATE {return $\bmu, \btheta^v$}
\ENDFUNCTION \
\REQUIRE {Network $\bPhi$, $\bp \in \bbR^{N_P(3+D_P)}, N, n_M, K \in \mathbb{N}^+, \sigma_1, \beta_1 \in \bbR^+$}
\STATE $\bmu \gets \mathbf{0}, \rho \gets 1, \btheta^v \gets [\frac{1}{K}]_{n_M \times K}$
\FOR{$i = 1$ to $N$}
\STATE $t \gets \frac{i-1}{n}$
\STATE $\hat{\bx}, \hat{\bv} \gets p_O(\bmu, \btheta^v, \bp, t)$
\STATE $\bmu, \btheta^v \gets \text{update}(\hat\bx, \hat\bv, \sigma_1, \beta_1, t)$
\ENDFOR
\STATE $\hat{\bx}, p_O^v(\hat{\bv}|\btheta^v, \bp; 1) \gets p_O(\bmu, \btheta^v, \bp, 1)$
\STATE $\hat{\bv} \sim p_O^v(\hat{\bv}|\btheta^v, \bp; 1)$
\STATE return $[\hat{\bx}, \hat{\bv}]$
\end{algorithmic}
\end{algorithm}

\section{Proof of SE-(3) Invariant Objective and SE-(3) Equivariant Sampling Process}\label{sec:se3_inv}

Density estimation and distribution learning on the 3D molecules should take translational and rotational invariance of the protein-ligand complex into consideration, \aka, the Special Euclidean group (SE-(3)) in 3D space \cite{satorras2021n, xu2021geodiff, hoogeboom2022equivariant}. 
Denote $T_g$ as the group of SE-(3) transformation, $T_g(\bx) = \bR\bx + \bb$, where $\bR \in \bbR^{3 \times 3}$ is the rotation matrix, and $\bb \in \bbR^3$ is the translation vector.

Following \citet{guan_3d_2023}, we move the center of protein to zero, \ie, $\tilde{\bp} = [\tilde{\bx}_P, \bv_P], \tilde{\bx}_P = \bQ\bx_P$, $\bQ = \mathbf{I}_3 \otimes (\mathbf{I}_{N_P} - \frac{1}{N_P}\mathbf{1}_{N_P}\mathbf{1}_{N_P}^\top)$ and shift molecule $\bm$ and variable $\bmu$ the same way. In another word, $\tilde\bp, \tilde\bm, \tilde\bmu$ are only defined with zero gravity of $\bp$, namely zero Center-of-Mass (CoM) space. 
That is, for any $T_g$ applied to the protein and molecule/variable complex, we always have $T_g(\tilde\bm, \tilde\bp) = \bR(\tilde{\bm}, \tilde{\bp}), T_g(\tilde\bmu, \tilde\bp) = \bR(\tilde{\bmu}, \tilde{\bp})$. Thus translational invariance is naturally satisfied on $\tilde\bp, \tilde\bm, \tilde\bmu$ by definition. For convenience, in the following discussion, we mention $\tilde\bp, \tilde\bm, \tilde\bmu$ as $\bp, \bm, \bmu$.

Then we slightly rewrite the key steps in Algorithm~\ref{algo:dloss}, and for convenience we omit $\btheta^v$, $\bv_M$, $\bv_P$ and $t$ since these variables are not in the 3D space.
\begin{align}
\bmu & = \gamma \bx_M + \gamma(1-\gamma)\beps \nonumber \\
\hat{\bx} & = \bPhi(\bmu, \bx_P) \nonumber \\
L_x^n(\bx_M, \bx_P) & = \mathrm{const} \Vert \bx_M - \hat{\bx} \Vert^2 \nonumber
\end{align}

Since $\epsilon$ is sampled from isotropic Gaussian, $\epsilon = \bR\epsilon'$ is from the same distribution. If we apply $T_g$ to the $\bmu, \bx_P$, since $\bPhi(\bmu, \bx_p)$ is an SE-(3) equivariant graph network, the output of network $\bPhi$ will be
\begin{align}
\bPhi(T_g(\bmu, \bx_P)) & = \bPhi(\bR(\bmu, \bx_P)) = \bR(\bPhi(\bmu, \bx_P)) = \bR(\hat\bx) = T_g(\hat\bx)
\end{align}

Thus the loss of $T_g$-transformed complex will be
\begin{align}
\Vert T_g(\bx_M) - T_g(\hat{\bx}) \Vert^2 & = \Vert \bR\bx_M + \bb - \bR\hat\bx - \bb \Vert^2 = \Vert \bR\bx_M - \bR\hat\bx \Vert^2 \nonumber \\
& = (\bx_M - \hat\bx)^\top \bR^\top \bR (\bx_M - \hat\bx) = (\bx_M - \hat\bx)^\top (\bx_M - \hat\bx) = \Vert \bx_M -\hat\bx \Vert^2
\end{align}
Thus the objective is SE-(3) invariant to $\bp, \bm, \bmu$.

\mycomment{
To meet the condition $\forall \bR \in \bbR^{3 \times 3}, \bt \in \bbR^3, \bR^\top \bR = \mathbf{I}, s.t., p_\theta(\bm | \bp) = p_\theta([\bR \bx_M + \bt, \bv_M] | [\bR \bx_P + \bt, \bv_P])$\kevin{exclude reflection?}, we have the following theorem:

\begin{theorem}
\label{thm:se3_trans}
With $\btheta$, $\by$, $\bx_M$, $\bx_P$ constrained in the zero Center of Mass (CoM) space~\cite{kohler2020equivariant}, the likelihood function $p_\theta$ is translational invariant.
\end{theorem}

\begin{theorem}
\label{thm:se3_rot}
When the following properties are satisfied, the likelihood function $p_\theta$ is rotational invariant:
\begin{align}
& \forall \bR, \bR^\top \bR = \mathbf{I}, s.t., \nonumber \\
& p_O(\bm|\btheta_0, \bp; t) = p_O([\bR\bx_M, \bv] | \btheta_0, [\bR\bx_P, \bv_P]; t), \nonumber \\
& p_O(\bm|\btheta_{}, \bp; t) = p_O([\bR\bx_M, \bv_M] | [\bR\btheta^x, \btheta^v], [\bR\bx_P, \bv_P]; t), \nonumber \\
& p_S(\by|\bm, \bp; \alpha) = p_S([\bR\by^x, \by^v]|[\bR\bx_M, \bv_M], [\bR\bx_P, \bv_P]; \alpha), \nonumber \\
& h([\bR\btheta^x, \btheta^v], [\bR\by^x, \by^v], \alpha) = [\bR h^x(\btheta, \by, \alpha), h^v(\btheta, \by, \alpha)]. \nonumber
\end{align}
\end{theorem}

\begin{proposition}
\label{prop:se3}
With the conditions in Th.~\ref{thm:se3_trans} and Th.~\ref{thm:se3_rot} satisfied, the variational lower bound in Eq.~\ref{eq:objective} is SE-(3) invariant.
\end{proposition}

We leave the proof of Th.~\ref{thm:se3_trans},~\ref{thm:se3_rot} and Prop.~\ref{prop:se3} in Appendix~\ref{}.
}

Before discuss the sampling process in Algorithm~\ref{algo:sample}, we slightly rewrite the key steps, and for convenience, we omit $\btheta^v, \bv_M, \bv_P$ and $t$ since these variables are not in the 3D space.
\begin{align}
\bmu_0 & = \mathbf{0}, \rho_0 = 1 \nonumber \\
\bx_{i} & = \bPhi(\bmu_{i-1}, \bx_P) \nonumber \\
\by^x_{i} & = \bx_{i} + \alpha_i\beps \nonumber \\
\rho_i & = \rho_{i-1} + \alpha_i \nonumber \\
\bmu_i & = \frac{\rho_{i-1} \bmu_{i-1} + \alpha_i \by^x_i}{\rho_{i}}.  \nonumber
\end{align}

At step 1, since $\bPhi(\bmu, \bx_P)$ is an SE-(3) equivariant graph network,
\begin{align}
T_g(\bx_1) & = T_g(\bPhi(\bmu_0, \bx_P)) = \bPhi(T_g(\bmu_0, \bx_P)) = \bPhi(\bmu_0, T_g(\bx_P)),
\end{align}
thus $\bx_1$ is SE-(3) Equivariant to $\bmu_0, \bx_P$.

For every step $i$, if we assume $T_g(\bx_{i}) = T_g(\bPhi(\bmu_{i-1}, \bx_P)) = \bPhi(T_g(\bmu_{i-1}, \bx_P))$, \ie, $\bx_{i}$ is SE-(3) equivariant to $\bmu_{i-1}, \bx_P$, we have (1) $\by_i^x$ is SE-(3) equivariant to $\bx_i$ since this update is simple addition and $\beps$ from isotropic Gaussian, (2) $\bmu_i$ is SE-(3) equivariant to $\bmu_{i-1}, \by_i^x$, thus $\bmu_i$ is SE-(3) equivariant to $\bx_i, \bx_P$. And $\bx_{i+1}$ is is SE-(3) equivariant to $\bmu_i, \bx_P$, thus $\bx_{i+1}$ is SE-(3) equivariant to $\bx_i, \bx_P$.

With mathematical induction, we have $\bx_i$ SE-(3) equivariant to $\bmu_0, \bx_P$, thus the final sample $\bx_N$ is also SE-(3) equivariant to $\bmu_0, \bx_P$. The sampling process in Algorithm~\ref{algo:sample} is SE-(3) equivariant to $\bx_P$.



\section{Implementation Details}
\subsection{Parameterization with SE-(3) Equivariant Network}\label{sec:se3net}
We model the interaction between ligand molecule atoms and protein pocket atoms with an SE-(3) equivariant network, PosNet3D \cite{guan2021energy}, as our backbone $\bPhi$ in Eq.~\ref{eq:p_o}.

A protein-molecule graph is firstly constructed through $k$-nearest neighbor search of the atom coordinates, $G = \langle V, E \rangle$. For each layer, 
the atom hidden states $\bh^l$ and coordinates $\bx^l$ are updated alternately as follows:
\begin{align}\footnotesize
\bh_i^{l+1} & = \bh_i^l + \sum_{j \in N_G(i)} \phi_{h}\Big(d_{ij}, \mathbf{h}_i^l, \mathbf{h}_j^l, \be_{ij}, t\Big) \\
\Delta \mathbf{x}_i & = \sum_{j \in N_G(i)} \Big(\bx_j^l - \bx_i^l\Big) \phi_x\Big(d_{ij}, \mathbf{h}_i^{l+1}, \mathbf{h}_j^{l+1}, \be_{ij}, t\Big) \nonumber \\
\bx_i^{l+1} & = \bx_i^l + \Delta \mathbf{x}_i \cdot 1_{mol}
\end{align}
where $N_G(i)$ denotes the neighborhood of atom $i$ in $G$, $(\bh_i, \bx_i)$ and $(\bh_j, \bx_j)$ denote atom $i$ and $j$, $d^l_{ij}$ is the euclidean distance between atoms $i$ and $j$, $\be_{ij}$ indicates the connection is between protein atoms, ligand atoms, or protein atom and ligand atom, $1_{mol}$ is to indicate only ligand atoms are updated, $\phi_h$ and $\phi_x$ are attention blocks which take $\bh_i^l$ as query and $[\bh_i^l, \bh_j^l, \be_{ij}]$ as keys and values. 

For the first layer, $\bx^0 = [\bmu, \bx_P]$, $\bh^0 = \text{linear}(\btheta^v, \bv_P, t)$. 
For the last layer, $\bPhi$ directly outputs an estimation $\hat{\bx} = \bPhi^x$. And for discrete variable $\bv^{(d)}$, $\bPhi$ takes softmax over the network output as distribution $\hat{\bv}^{(d)} = \text{softmax}\Big((\bPhi^v)^{(d)}\Big)$.

\subsection{Featurization}
For each protein atom, we represent it by several features, including a one-hot element indicator (H, C, N, O, S, Se) to identify the element type, a one-hot amino acid type indicator (20 dimension) to indicate the amino acid type, a one-dimension flag to indicate if the atom is a backbone atom, and a one-hot arm/scaffold region indicator to determine if the atom belongs to an arm or scaffold region based on its distance from the arm prior center. 

The features for the ligand atom include a one-hot element indicator (C, N, O, F, P, S, Cl) to represent the element type, and a one-hot arm/scaffold indicator to differentiate between aromatic and non-aromatic atoms. 

Two graphs are dynamically built for message passing in the protein-ligand complex, a k-nearest neighbors graph for ligand atoms and protein atoms, and a fully-connected graph for ligand atoms. The edge features in the k-nearest neighbors graph are the outer products of distance embedding (obtained by expanding the distance using radial basis functions) and edge type (a 4-dim one-hot vector indicating the type of edge). The ligand graph represents ligand bonds with a one-hot bond type vector (non-bond, single, double, triple, aromatic).

\subsection{Model Hyperparameters}
For the SE-(3) equivariant network, we experiment with kNN graphs with 32-nearest neighbor search to construct graph, 9 layers with hidden dimension of 128, 16 head attention, ReLU activation with Layer Normalization \cite{ba2016layer}.

For the noise schedules, we use $\beta_1 = 1.5$ for atom types, $\sigma_1 = 0.03$ for atom coordinates, and train the model with discrete time loss of 1000 training steps.

For training, we use Adam optimizer with learning rate 0.005, batch size of 8, and exponential moving average of model parameters with a factor of 0.999. The training will converge within 15 epochs on a single RTX 3090, taking around 24 hours.  
For sampling, we take 100 sample steps with noise-reduced sampling strategy.


\section{Additional Experimental Results}\label{sec:exp_detail}
\subsection{Full Evaluation Results}
\paragraph{Binding Interaction}
In addition to our main results in Table~\ref{tab:main_results}, we provide evaluation in terms of key interactions in Table~\ref{tab:key_interaction}, i.e., No. Hydrogen Bond Donors (HB Donors), No. Hydrogen Bond Acceptors (HB Acceptors), van-der Waals contacts (vdWs) and Hydrophobic interactions as described in PoseCheck \citep{harris2023benchmarking}. It can be seen that under different molecule sizes, ours is consistently the best in forming hydrogen bonds, indicating that \ours~finely captures key protein-ligand interactions. 

\begin{table}[htbp]
\centering
\caption{Number of key interactions for SBDD models under different molecule sizes. Top-1 values highlighted in \textbf{bold} text.}
\label{tab:key_interaction}
\resizebox{\linewidth}{!}{

\begin{tabular}{@{}clccccc@{}}
\toprule
                                &            & HB Donors (Avg.) & HB Acceptor (Avg.) & vdWs (Avg.)   & Hydrophobic (Avg.) & Molecule Size \\ \midrule
                                & Reference  & 0.87             & 1.42               & 6.61          & 5.06               & 22.8          \\ \midrule
\multirow{4}{*}{Smaller-size}     & AR         & 0.51             & 0.90               & 5.54          & 3.78               & 17.7          \\
                                & Pocket2Mol & 0.32             & 0.63               & 5.25          & \textbf{4.53}      & 17.7          \\
                                & FLAG       & 0.28             & 0.30               & 5.85          & 3.76               & 16.7          \\
                                & Ours-small & \textbf{0.62}    & \textbf{1.09}      & \textbf{6.24} & 4.42               & 17.8          \\ \midrule
\multirow{3}{*}{Reference-size} & TargetDiff & 0.63             & 0.98               & \textbf{7.92} & \textbf{5.43}      & 24.2          \\
                                & Decomp-R   & 0.56             & 0.99               & 6.70          & 4.37               & 21.2          \\
                                & Ours       & \textbf{0.71}    & \textbf{1.25}      & 7.38 & 5.07               & 22.7          \\ \midrule
\multirow{2}{*}{Larger-size}     & Decomp-O   & 0.52             & 0.87               & \textbf{9.14} & \textbf{6.84}      & 29.4          \\
                                & Ours-large & \textbf{0.75}    & \textbf{1.38}      & 8.64          & 6.07               & 29.4          \\ \bottomrule 
\end{tabular}
}
\end{table}

\begin{figure*}[htbp]
\centering
\includegraphics[width=\textwidth]{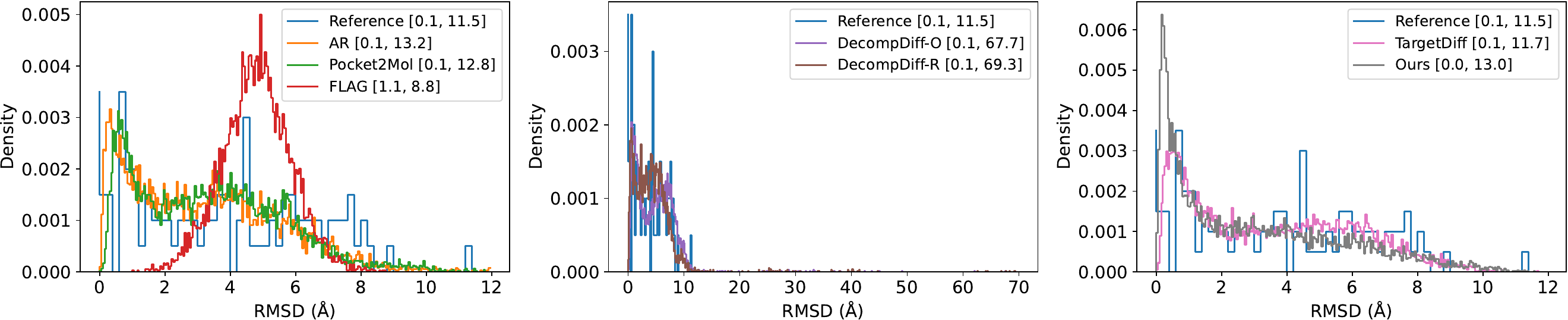}
\caption{RMSD distribution of generated molecules compared with reference molecules. \emph{Note:} values in [] are RMSD ranges for each model, which generally lie in [0, 13] except for significant outliers of DecompDiff.}
\label{fig:rmsd}
\end{figure*}

\begin{figure*}[h]
\centering
\subfigure[C-C Bond]{\includegraphics[width=\textwidth]{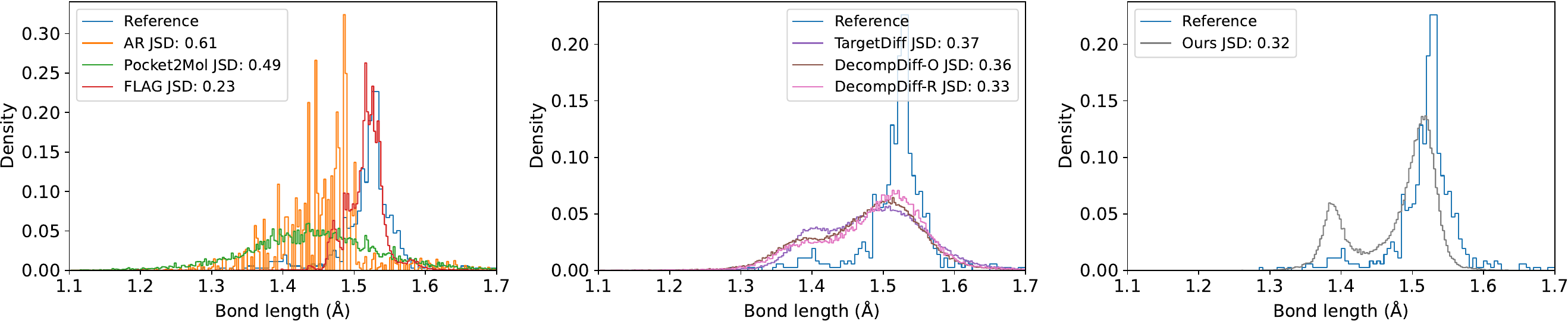}}
\subfigure[C:C Bond]{\includegraphics[width=\textwidth]{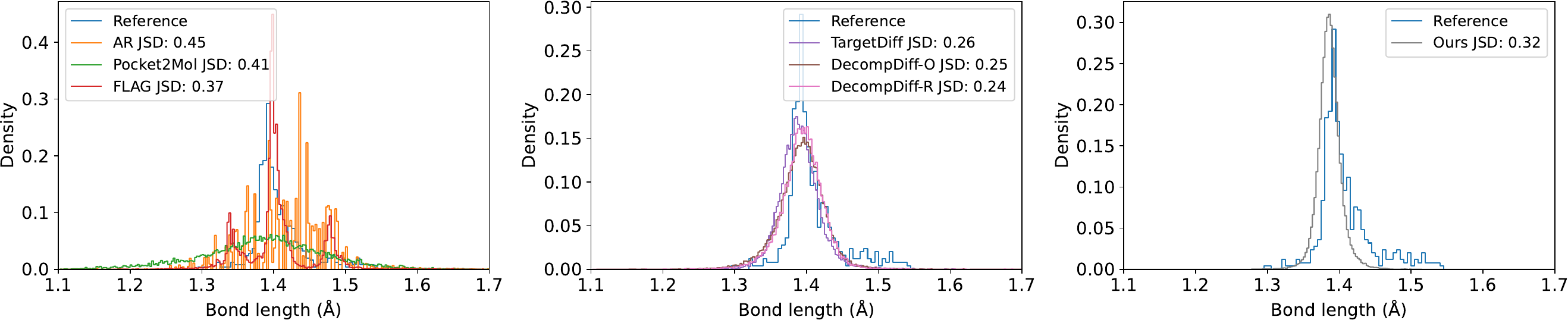}}
\subfigure[C-O Bond]{\includegraphics[width=\textwidth]{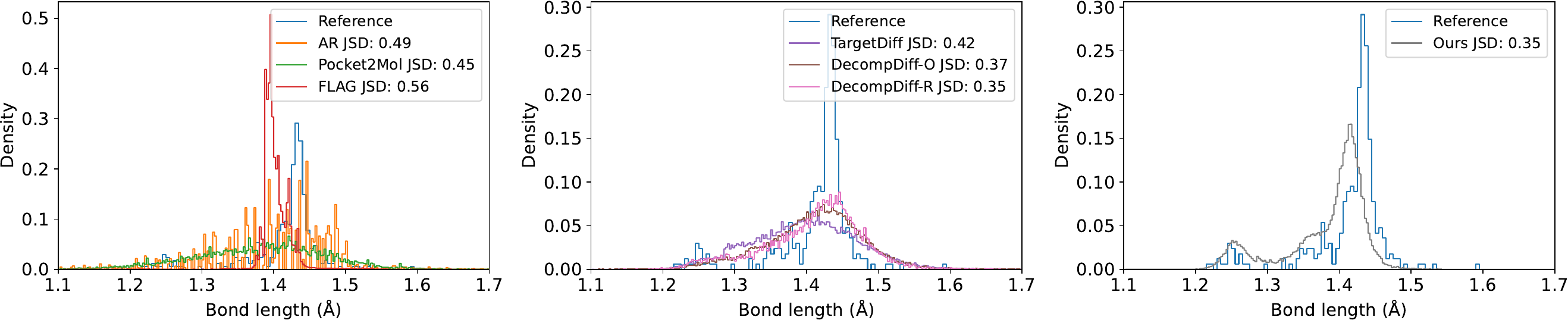}}
\subfigure[C-N Bond]{\includegraphics[width=\textwidth]{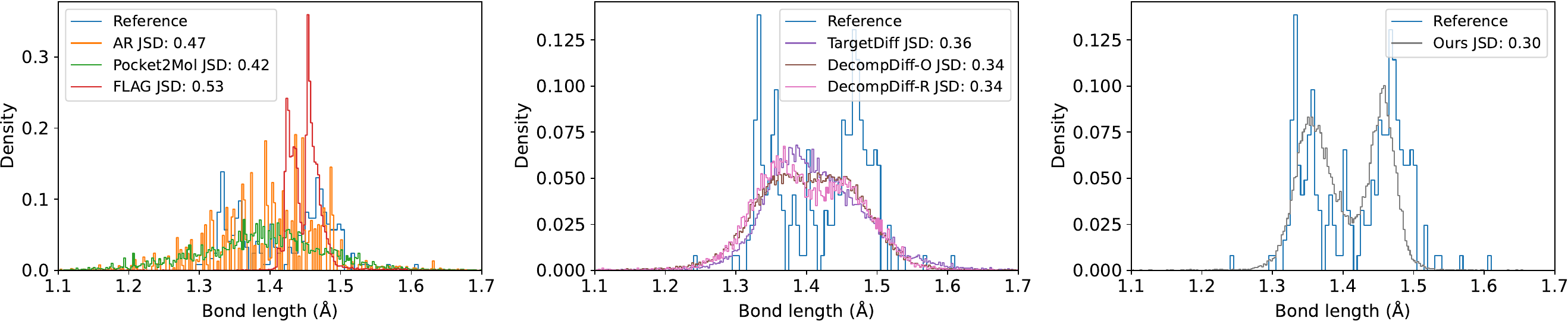}}
\subfigure[C:N Bond]{\includegraphics[width=\textwidth]{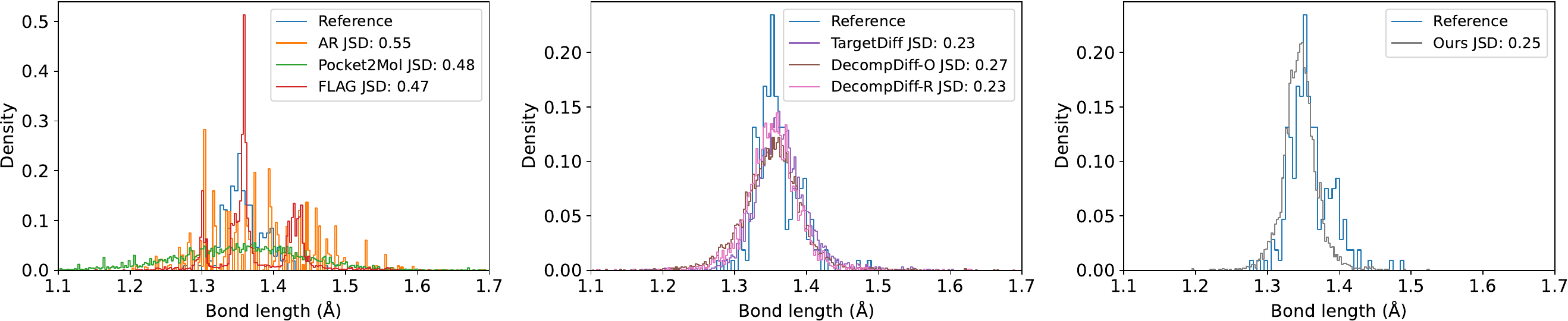}}
\caption{Bond length distribution of generated molecules compared with reference molecules.}
\label{fig:other_len}
\end{figure*}

\begin{figure*}[h]
\centering
\subfigure[C-C-C Bond Angle]{\includegraphics[width=\textwidth]{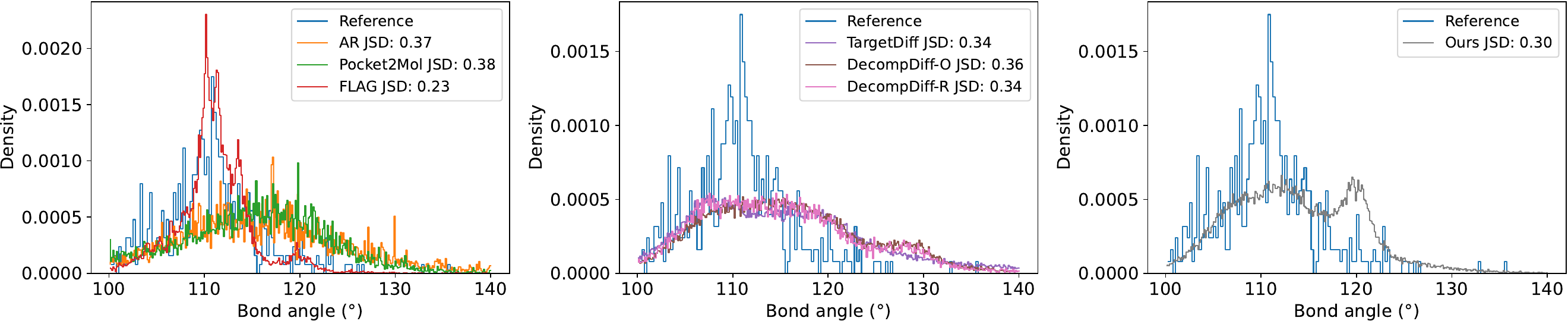}}
\subfigure[C:C:C Bond Angle]{\includegraphics[width=\textwidth]{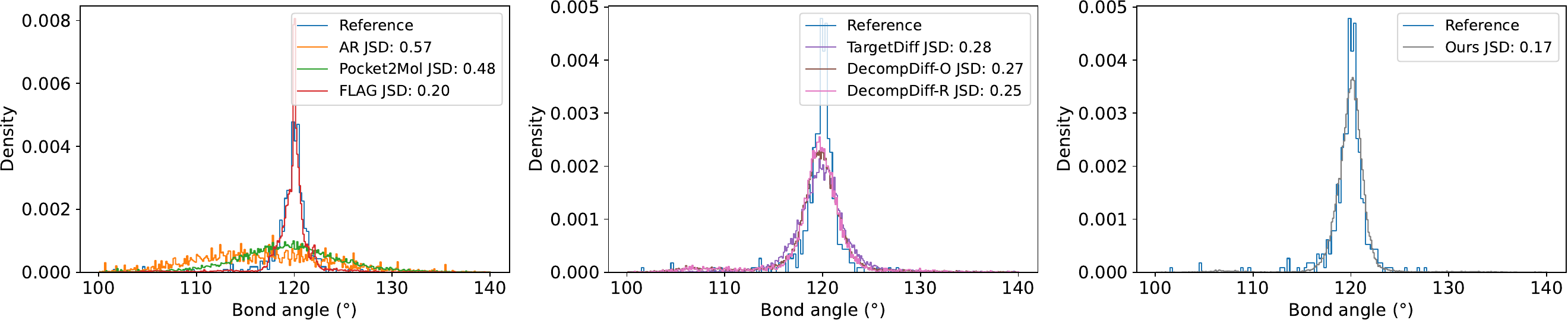}}
\subfigure[C-C-O Bond Angle]{\includegraphics[width=\textwidth]{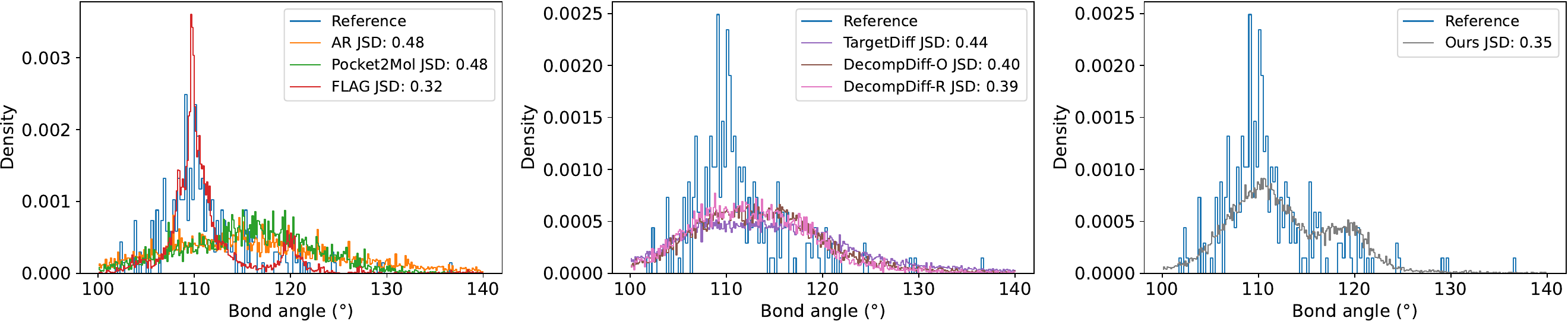}}
\caption{Bond angle distribution of generated molecules compared with reference molecules.}
\label{fig:other_angle}
\end{figure*}

\begin{figure*}[ht]
\centering
\subfigure[C-C-C-C Torsion Angle]{\includegraphics[width=\textwidth]{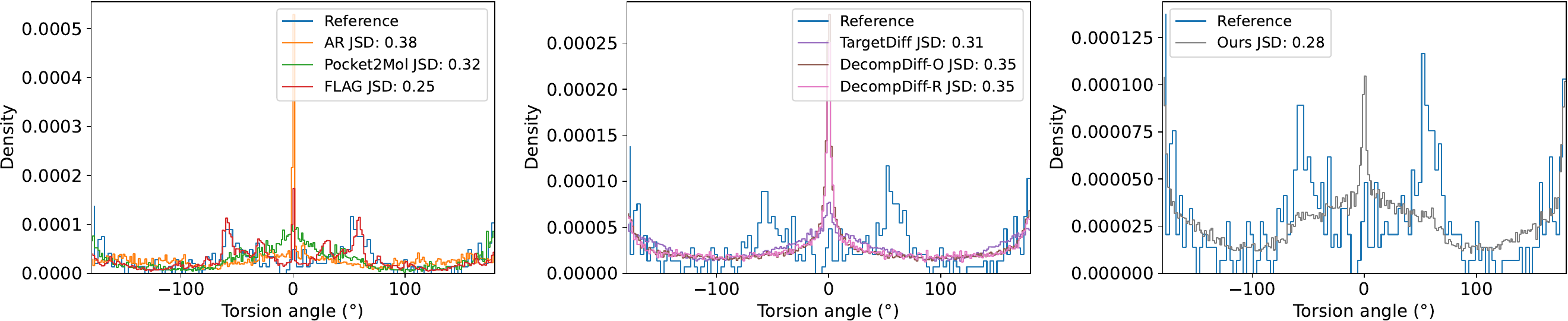}}
\subfigure[C:C:C:C Torsion Angle]{\includegraphics[width=\textwidth]{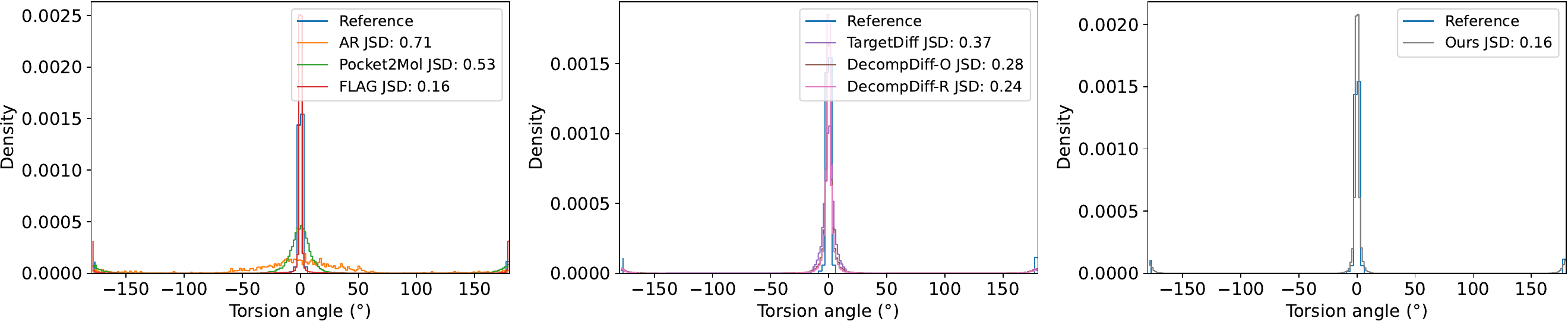}}
\caption{Torsion angle distribution of generated molecules compared with reference molecules.}
\label{fig:other_torsion}
\end{figure*}

\paragraph{Comformation Stability}
Besides the substructural analyis in Table~\ref{tab:posecheck_results}, we present the full evaluation results for bond length, bond angle and torsion angle distributions of different types are presented in Table~\ref{tab:bond_len}, \ref{tab:bond_angle}, and \ref{tab:torsion_angle}, and further visualize the distributions in Fig.~\ref{fig:other_len}, \ref{fig:other_angle}, and \ref{fig:other_torsion}, in order to look into details of 3D local structures.

As shown in Table~\ref{tab:torsion_angle}, FLAG \citep{zhang_molecule_2023} displays the best torsion angle distribution, owing to its fragment-based sampling strategy. However, FLAG has been observed to generate 0.2\% severe outlier molecules that could not be parsed in JSD calculation, casting doubt on its local structures associated with fragment linkers. 

Another perspective is given in Fig.~\ref{fig:all_len}, where we find that atom-based autoregressive models can hardly distinguish different bond types, yet diffusion models and our model show clear pattern similar to reference, which demonstrates another drawback of autoregressive models for SBDD. Since FLAG is a fragment-based autoregressive model, it inherits natural bonds by its nature. AR is a voxel-based model, thus its curves are not smooth as expected.

Next, we move on for a closer inspection into conformation stability of binding complex.
According to Table~\ref{tab:main_results}, our model has the best RMSD of binding poses, which means 46\% of our generated molecules already have accurate docking pose even without force field optimization or redocking, rendering our results more reliable for drug discovery. A notable phenomenon is the extremely low rate of FLAG, thus we investigate the overall distributions in Fig.~\ref{fig:rmsd}. From this figure, we find AR \citep{luo_3d_2022}, Pocket2Mol \citep{peng_pocket2mol_2022}, TargetDiff and our model have closer distribution and value ranges to reference, yet FLAG has an obvious mode around 5, which explains the extremely low RMSD of FLAG.

For the reason why our model has better RMSD than reference, a possible explanation is that in CrossDocked \citep{francoeur2020three} training set, 52.4\% poses are obtained via Vina Dock, while in the test set, only 37\% molecules undergo redocking. This ratio accounts for the reference contains 34\% RMSD below 2\r{A}, and ours (46.1\%) is approaching the limit of training set (52.4\%).

Besides RMSD, we also report steric clashes \cite{harris2023benchmarking} in protein-ligand complex conformation. As shown in Table~\ref{tab:posecheck_results}, we achieve considerably fewer steric clashes. 
It could be seen that DecompDiff does surpass TargetDiff with the help of better priors, but it also suffers from the distribution shift of molecular size (29.4, larger than reference 22.8) introduced by manually chosen priors.



\begin{figure*}[ht]
    \centering
    \includegraphics[width=\linewidth]{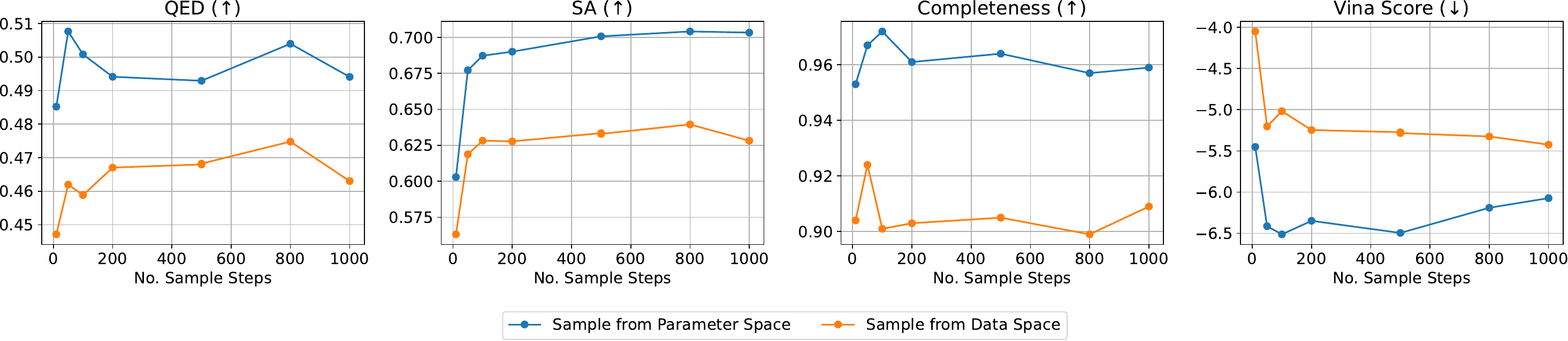}
    \caption{Ablation study on our proposed noise reduced sampling strategy under different sampling steps. Higher QED, SA, Completeness and lower Vina Score indicate better performance. The model is trained with 1000 steps.}
    \label{fig:ablation}
\end{figure*}

\subsection{Ablation Studies}\label{subsec:app_ablation}

As described in Fig.~\ref{fig:sample_efficiency}, our method samples at significantly faster speed.
The reason of our considerable speed up mainly contributes to our improved sampling strategy from the parameter space, discussed in Sec.~\ref{subsec:train_sample}. For this reason, TargetDiff and DecompDiff requires 1000 steps for sampling, yet our model requires much fewer sampling steps to achieve comparable results. 
Therefore, we conduct ablation studies on sampling strategy and sampling steps so as to validate our design.

We test different sampling strategies with different steps for the same checkpoint, and sample 10 molecules each for 100 test proteins. We plot the curves of QED, SA, Completeness $(\uparrow)$ and Vina Score $(\downarrow)$ in Figure~\ref{fig:ablation}.
As the sampling step increases to training steps, we found the original sampling strategy exhibits first enhanced then slightly decreased sample quality, possibly because the update of parameters is smoothed / oversmoothed by finer partitioned noise factor $\alpha$, whereas the noise reduced strategy displays this tendency far earlier and generates the best quality of molecules with fewer sampling steps, indicating its high efficiency. Considering the overall sample quality, we decide to use 100 sampling steps for our model.

\begin{table}[htbp]
\caption{Jensen-Shannon divergence of top-8 frequent bond length distributions between the reference and the generated molecules ($\downarrow$ is better). No connected line, “=”, and “:” represent single, double, and aromatic bonds. Top 2 results are highlighted with \textbf{bold text} and \ul{underlined text}, respectively.}
\label{tab:bond_len}
\vskip 0.1in
\centering
\begin{tabular}{@{}c|ccc|cccc@{}}
\toprule
Bond & AR    & Pocket2Mol & FLAG           & TargetDiff     & Decomp-O & Decomp-R & Ours           \\ \midrule
CC        & 0.610 & 0.494      & \textbf{0.231} & 0.369          & 0.359             & 0.326          & \ul{0.290}    \\
C:C       & 0.450 & 0.414      & 0.366          & 0.263          & \ul{0.251}       & \textbf{0.238} & 0.330          \\
CO        & 0.490 & 0.452      & 0.556          & 0.421          & 0.375             & \ul{0.346}    & \textbf{0.339} \\
CN        & 0.472 & 0.422      & 0.529          & 0.362          & \ul{0.342}       & 0.337          & \textbf{0.292} \\
C:N       & 0.551 & 0.484      & 0.470          & \textbf{0.235} & 0.269             & \textbf{0.235} & \ul{0.242}    \\
OP        & 0.676 & 0.523      & 0.690          & 0.441          & \ul{0.435}       & 0.450          & \textbf{0.347} \\
C=O       & 0.556 & 0.510      & 0.638          & 0.461          & \ul{0.368}       & 0.391          & \textbf{0.342} \\
O=P       & 0.626 & 0.581      & 0.609          & 0.506          & 0.472             & \ul{0.458}    & \textbf{0.369} \\ \midrule
Avg.      & 0.554 & 0.485      & 0.511 & 0.382      & 0.359             & \ul{0.348}          & \textbf{0.319} \\ \bottomrule
\end{tabular}
\vskip -0.1in
\end{table}

\begin{table}[htbp]
\caption{Jensen-Shannon divergence of top-8 frequent bond angle distributions between the reference and the generated molecules ($\downarrow$ is better). Top 2 results are highlighted with \textbf{bold text} and \ul{underlined text}.}
\label{tab:bond_angle}
\vskip 0.1in
\centering
\begin{tabular}{@{}c|ccc|cccc@{}}
\toprule
Angle & AR    & Pocket2Mol     & FLAG           & TargetDiff & Decomp-O & Decomp-R & Ours           \\ \midrule
CCC   & 0.372 & 0.380          & \textbf{0.231} & 0.345      & 0.358             & 0.337          & \ul{0.280}    \\
C:C:C & 0.572 & 0.480          & \ul{0.199}    & 0.283      & 0.266             & 0.255          & \textbf{0.172} \\
CCO   & 0.477 & 0.475          & \textbf{0.318} & 0.440      & 0.403             & 0.394          & \ul{0.319}    \\
C:C:N & 0.537 & 0.506          & 0.465          & 0.454      & \textbf{0.429}    & \textbf{0.429} & \ul{0.446}    \\
CCN   & 0.447 & 0.443          & \ul{0.388}    & 0.437      & 0.404             & 0.419          & \textbf{0.377} \\
CNC   & 0.535 & \textbf{0.498} & 0.510          & 0.521      & \textbf{0.498}    & 0.504          & \ul{0.499}    \\
COC   & 0.496 & 0.494          & 0.607          & 0.502      & \ul{0.484}       & 0.492          & \textbf{0.460} \\
C:N:C & 0.619 & 0.580          & 0.526          & 0.495      & \ul{0.473}       & \textbf{0.462} & 0.475          \\ \midrule
Avg.      & 0.507 & 0.482      & \ul{0.406} & 0.435      & 0.414             & 0.412          & \textbf{0.379} \\ \bottomrule
\end{tabular}
\end{table}

\begin{table}[htbp]
\caption{Jensen-Shannon divergence of top-8 frequent torsion angle distributions between the reference and the generated molecules ($\downarrow$ is better). Top 2 results are highlighted with \textbf{bold text} and \ul{underlined text}.}
\label{tab:torsion_angle}
\vskip 0.1in
\centering
\begin{tabular}{@{}c|cc|cccc@{}}
\toprule
Torsion Angle & AR          & Pocket2Mol & TargetDiff     & Decomp-O & Decomp-R & Ours           \\ \midrule
CCCC          & 0.378       & 0.320      & \ul{0.312}    & 0.349             & 0.348          & \textbf{0.286} \\
C:C:C:C       & 0.704       & 0.514      & 0.348          & 0.264             & \ul{0.230}    & \textbf{0.130} \\
CCOC          & 0.419       & 0.401      & \textbf{0.390} & 0.392             & \ul{0.391}    & 0.393          \\
CCCO          & 0.431       & 0.405      & 0.403          & \ul{0.402}       & 0.403          & \textbf{0.396} \\
CCNC          & 0.430       & 0.437      & 0.423          & 0.403             & \textbf{0.398} & \ul{0.401}    \\
C:C:N:C       & 0.664       & 0.504      & 0.386          & 0.285             & \textbf{0.197} & \ul{0.212}    \\
C:C:C:N       & 0.663       & 0.512      & 0.441          & 0.388             & \ul{0.316}    & \textbf{0.281} \\
C:N:C:N       & 0.742       & 0.535      & 0.476          & 0.366             & \textbf{0.247} & \ul{0.303}    \\
CCCN          & \ul{0.495} & 0.549      & 0.512          & 0.501             & 0.525          & \textbf{0.493} \\ \midrule
Avg.      & 0.547 & 0.464      & 0.410 & 0.372      & \ul{0.340}             & \textbf{0.322} \\ \bottomrule
\end{tabular}
\end{table}


\begin{figure*}[htbp]
    \centering
    \includegraphics[width=\linewidth]{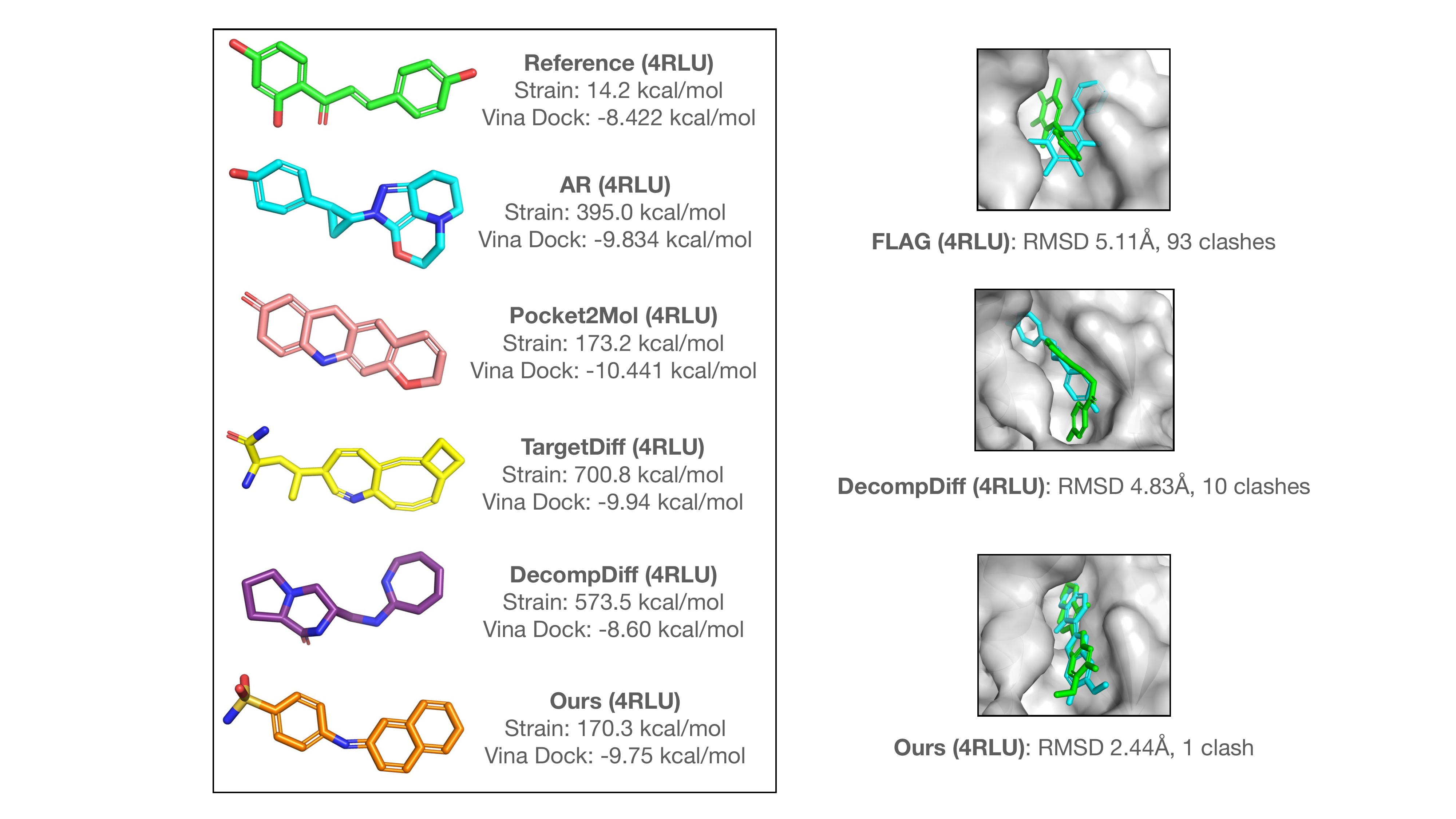}
    \caption{Visualization of molecules for a randomly chosen test protein (PDB ID: 4RLU), representative in the median values of strain energy (Strain) and RMSD.}
    \label{fig:visualize_mol}
\end{figure*}



\end{document}